\documentclass{IEEEtran}
\usepackage[utf8]{inputenc}
\usepackage[T1]{fontenc}
\usepackage{amsmath,bm}
\usepackage{amsfonts}
\usepackage{amssymb}
\usepackage{graphics}
\usepackage{epsfig}
\usepackage{subfigure}
\usepackage{caption}
\usepackage{amsthm}
\usepackage{algorithm}
\usepackage{setspace}
\usepackage{enumitem,kantlipsum}

\usepackage[noend]{algpseudocode}
\usepackage{mathrsfs}
\usepackage{color}
\usepackage{cite}
\usepackage{dblfloatfix}

\usepackage{tikz}
\usetikzlibrary{shapes,arrows}

\graphicspath{{figure/}}

\newtheorem{definition}{Definition}
\newtheorem{Theorem}{Theorem}
\newtheorem{Lemma}{Lemma}
\newtheorem{Problem}{Problem}
\newtheorem{Corollary}{Corollary}
\newtheorem{Remark}{Remark}
\newtheorem{Assumption}{Assumption}

\makeatletter
\def\BState{\State\hskip-\ALG@thistlm}
\makeatother

\usepackage{tikz}
	\usetikzlibrary{shapes,arrows}
	\tikzstyle{frame} = [draw, -latex]
	\tikzstyle{line} = [draw]
	\tikzstyle{line2} = [draw, dashdotted]
	\tikzstyle{line3} = [draw, dashed]
	\tikzstyle{line3UD} = [draw, dashed]
	\tikzstyle{place} = [circle, draw=black, fill=white, thick, inner sep=2pt, minimum size=1mm]
	\tikzstyle{place2} = [circle, draw=black, fill=black, thick, inner sep=2pt, minimum size=1mm]
	\tikzstyle{placeRed} = [circle, draw=red, fill=red, thick, inner sep=2pt, minimum size=1mm]
	\tikzstyle{vertex} = [circle, draw=black, fill=black, thick, inner sep=2pt, minimum size=1mm]
\usepackage{tkz-euclide}
\usepackage[switch,pagewise]{lineno}
\newcommand*\patchAmsMathEnvironmentForLineno[1]{%
\expandafter\let\csname old#1\expandafter\endcsname\csname #1\endcsname
\expandafter\let\csname oldend#1\expandafter\endcsname\csname end#1\endcsname
\renewenvironment{#1}%
{\linenomath\csname old#1\endcsname}%
{\csname oldend#1\endcsname\endlinenomath}}%
\newcommand*\patchBothAmsMathEnvironmentsForLineno[1]{%
\patchAmsMathEnvironmentForLineno{#1}%
\patchAmsMathEnvironmentForLineno{#1*}}%
\AtBeginDocument{%
\patchBothAmsMathEnvironmentsForLineno{equation}%
\patchBothAmsMathEnvironmentsForLineno{align}%
\patchBothAmsMathEnvironmentsForLineno{flalign}%
\patchBothAmsMathEnvironmentsForLineno{alignat}%
\patchBothAmsMathEnvironmentsForLineno{gather}%
\patchBothAmsMathEnvironmentsForLineno{multline}%
}
\begin{document}
\title{Distributed Bearing-based Formation Control and Network Localization with Exogenous Disturbances}

\author{Yoo-Bin Bae, Seong-Ho Kwon, Young-Hun Lim, and Hyo-Sung Ahn
%\thanks{This work was supported by the National Research Foundation (NRF) of Korea under the grant NRF-2017R1A2B3007034.}
\thanks{Y.-B. Bae, S.-H. Kwon, and H.-S. Ahn are with Distributed Control \& Autonomous Systems Lab. (DCASL), School of Mechanical Engineering, Gwangju Institute of Science and Technology (GIST), Gwangju 500-712, Republic of Korea. E-mail: {\tt\small \{bub0418, seongho, hyosung\}@gist.ac.kr.}}
\thanks{Y.-H Lim is with the Department of Electronic Engineering, Gyeongnam National University of Science and Technology(GNTECH), 33 Dongjin-ro, Jinju, 52725, Republic of Korea. E-mail:  {\tt\small yhunlim@gntech.ac.kr.}}
}

\maketitle

\begin{abstract}
This paper presents a generalized robust stability analysis for bearing-based formation control and network localization systems. For an undirected network, we provide a robust stability analysis in the presence of time-varying exogenous disturbances in arbitrary dimensional space. In addition, we compute the explicit upper-bound set of the bearing formation and network localization errors, which provides valuable information for a system design. 
\end{abstract}

\begin{IEEEkeywords}
Formation control, multi-agent systems, network localization, robustness.
\end{IEEEkeywords}

\section{Introduction}
\IEEEPARstart{M}{ulti-agent} coordination tasks have been studied recently due to their great potential for various applications in the real-world. Formation control and network localization of multi-agent systems are the two main coordination task problems. Referring to the paper \cite{oh2015survey}, depending on the measured and controlled variables, the existing formation control problems are categorized into a position-based, displacement-based, distance-based and bearing-based formation control. Likewise, depending on the measured variables, the existing network localization problems can be categorized into position-based, distance-based, and bearing-based network localization.
Latest research efforts have been focused on the coordination tasks in a distributed manner that uses less information on measurement and control since it has many advantages, e.g., cost-effective, simplicity of the sensor systems, in practical applications. In this aspect, recently, there has been much literature on the distributed bearing-based formation control \cite{bishop2011stabilization, zhao2015bearing, zhao2014distributed, schoof2014bearing, zhao2015bearingtac, trinh2018bearing, zhao2018revisit, tron2016bearing, zhao2019bearing} and network localization \cite{eren2007using, shames2012analysis, zhu2014distributed, zhao2016localizability} since the bearing sensing capability is a minimal requirement of agent compared to the other measurements. In the real-world, bearing measurements can be obtained by an on-board vision sensors \cite{montijano2016vision}. 

A recent issue on bearing-based formation control for the undirected network was how to design a bearing-only controller, which uses bearing-only information for formation control tasks. Early works proposed a bearing-only controller with auxiliary measurements \cite{bishop2011stabilization, zhao2015bearing}, or their works were confined to particular formations and plane \cite{zhao2014distributed, schoof2014bearing}. Afterward, \cite{zhao2015bearingtac} proposed a bearing-only controller that can ensure almost global convergence in arbitrary dimensional space. Also, \cite{tron2016bearing, zhao2018revisit, zhao2019bearing} studied a gradient-descent bearing-only controller that can handle a more realistic model, e.g., moving target formation. 
Likewise, there were recent advances in distributed bearing-based network localization systems. Unlike the previous works \cite{eren2007using, shames2012analysis, zhu2014distributed}, which are only applicable in two-dimensional space, \cite{zhao2016localizability} proposed distributed network localization protocol that can globally localize a localizable network in arbitrary dimensional space.

However, most of the previous works on bearing-based formation control and network localization are studied under ideal conditions. In the real-world, various sources of disturbances (e.g., sensor error, command error, device aging, and wind) exist, and these factors interrupt our mission objectives. Therefore, it is worth analyzing the correlation of the disturbances and system error for practical applications. Despite the importance of robust stability analysis, as far as the authors are concerned, there are only a few relevant studies on bearing-based formation control and network localization systems. To be specific, for the leader-follower bearing-only formation tracking control system, \cite{zhao2015bearingtac} considered constant input disturbances to the followers. They showed that the proposed controller compensates for the unknown disturbances. 
On the other hand, for the network localization system, \cite{shames2012analysis} assumed the existence of the low level bearing measurement errors in two-dimensional space. A bound of the network localization error was derived, and they proposed a method on how to choose anchor nodes to minimize the effect of measurement errors. And, \cite{zhao2016localizability} considered constant measurement errors on bearing and anchor's location. They gave a bound for total bearing measurement errors to ensure the stability of the localization system. Also, a bound for the network localization error was derived when the measurement errors on the anchor's location exist. In our earlier work \cite{bae2019leader}, a similar analysis to this paper was presented for a leader-fixed follower formation control system. However, the formation scale was assumed to be upper-bounded. We extend our earlier work by deleting such an assumption in Section \ref{sec33}.

Unlike the aforementioned previous works on robustness issues that have limitations, i.e., constant disturbances, specific source of disturbances, and two-dimensional space, this paper presents a generalized robust stability analysis for bearing-based formation control and network localization in arbitrary dimensional space. For an undirected network with the knowledge of common global coordinate frame, based on the conventional distributed bearing-based formation control and network localization protocols, we consider the time-varying generalized exogenous disturbances as an additive term on single-integrator dynamics. Then, we investigate the stability of the formation control and network localization systems. Also, the correlation between exogenous disturbances and system errors will be analyzed by way of computing the explicit upper-bound of the bearing formation and network localization errors. For instance, assuming that the system error tolerance is known, we can check whether the error is allowable or not by computing the upper-bound of bearing formation and network localization errors in practical tasks.

Consequently, the main contribution of this paper is to present a generalized robust stability analysis for bearing-based formation control and network localization systems with the time-varying exogenous disturbances. In addition, we compute the explicit upper-bound of the bearing formation and network localization errors to investigate the correlation between system factors including exogenous disturbances and errors, which gives useful information for system design. %Lastly, a new Lyapunov stability analysis that can handle moving formations with the exogenous disturbances for a leaderless formation control system is carried out as a foundation for robust stability analysis.

The rest of this paper is organized as follows. In Section \ref{sec2}, we introduce basic graph theory and background knowledge for infinitesimal bearing rigidity and network localizability. In Section \ref{sec3}, two bearing-based formation control systems: 1) leaderless, and 2) leader-fixed follower are analyzed. In Section \ref{sec4}, a bearing-based network localization system is analyzed. Section \ref{sec5} shows the simulation results. Finally, conclusions are drawn in Section \ref{sec6}.

\section{Preliminaries}\label{sec2}
\subsection{Basic graph theory}
The interaction topology of a multi-agent system in this paper is modeled by an undirected graph, represented by $\mathcal{G = (V, E)}$ with node set $\mathcal{V}$ and edge set $\mathcal{E}$, i.e., $|\mathcal{V}|=n$ and $|\mathcal{E}|=m$. For leader-fixed follower multi-agent system, the node set $\mathcal{V}$ consists of leader node set $\mathcal{V}_L$ and follower node set $\mathcal{V}_F$ where $\mathcal{V} = \mathcal{V}_L \cup \mathcal{V}_F$, i.e., $|\mathcal{V}_L|=n_l, |\mathcal{V}_F|=n_f, n = n_l + n_f$. Without loss of generality, we suppose that the first $n_l$ nodes are leaders and the remaining $n_f$ nodes are followers. The neighborhood set of arbitrary node $i$ is defined as $\mathcal{N}_i = \{j \in \mathcal{V} : \>(i,j)\in \mathcal{E} \}$. In $d$-dimensional space, the position vector of arbitrary node $i$ is denoted by $p_i \in \mathbb{R}^{d}$, and the configuration, which is the overall position vector set, is denoted by $p = [p_1^T, \ p_2^T,\ .\ .\ . \ ,p_n^T]^T \in \mathbb{R}^{nd}$. Then, a framework is defined as a pair of $\mathcal{G}$ and $p$, i.e., $(\mathcal{G}, p)$. %For leader-follower formation system, the configuration can be denoted by $p= [p_{l}^T, \ p_{f}^T]^T \in \mathbb{R}^{nd}$ where $p_l(t) = p_l^{*}$, which means that the leader agents are stationary to their desired positions $p_l^*$.%

For an undirected graph, an arbitrary orientation can be given for each edge. In this paper, for an edge $(i, j)$ where $i<j$, the node $i$ is called as the head node, and the node $j$ is called as the tail node. The incidence matrix $H = [h_{ki}] \in \mathbb{R}^{m \times n}$ for an undirected oriented graph is defined as $h_{ki} = -1$ (respectively, 1) if node $i$ is the head (respectively, tail) node and otherwise, $h_{ki}=0$ where $i, j$ vertices are connected by $k^{th}$ oriented edge. %We note that $\rank(H)=n-1$ for a connected graph \cite{mesbahi2010graph}. 
The relative position vector between nodes $i$ and $j$ is denoted by $z_{k_{ij}} = p_j - p_i$ (For notational convenience, we will use $z_k$ instead of $z_{k_{ij}}$, where $k \in \{1,2,\cdots,m \}$), and the relative position vector set is denoted by $z = [z_{1}^T,\ z_{2}^T, \ .\ .\ .\ ,z_{m}^T]^T \in \mathbb{R}^{md}$. We can write $z = \bar{H}p=(H \otimes I_d)p$, where $\otimes$ is the Kronecker product and $I_d$ is the $d \times d$ identity matrix. 
The relative bearing vector between nodes $i$ and $j$ is defined as 
\begin{equation*}\label{beardef}
g_{k}:=\frac{z_k}{||z_k||},
\end{equation*}
where $|| \cdot ||$ denotes the Euclidean norm for a vector or the spectral norm for a matrix. The relative bearing vector set (also called as the bearing function) is denoted by $g =[g_1^T, \ g_2^T,\ .\ .\ . \ ,g_m^T]^T \in \mathbb{R}^{md}$.
\subsection{Background knowledge}
We first introduce the notion of infinitesimal bearing rigidity.
% which will be useful for analysis of the leaderless formation control system (Section \ref{sec32}).
The bearing rigidity matrix is defined as the Jacobian of the bearing function \cite{zhao2015bearingtac}
\begin{equation*} \label{brigid}
R_b(p) := \frac{\partial{g}}{\partial{p}} \in \mathbb{R}^{md \times nd}.
\end{equation*}
If $p^{\prime} \in \mathrm{Null}(R_b(p))$, where $p^{\prime}$ is a variation of configuration, then $p^{\prime}$ is called an infinitesimal bearing motion. The bearing-preserving motions of the framework include translational and scaling motions. Then, an infinitesimal bearing motion is called trivial  if it only includes translational and scaling motions of the framework. 

We next show the definition and lemma for infinitesimal bearing rigidity of framework from the knowledge in \cite{zhao2015bearingtac}.

\begin{definition}\label{infbearing}
A framework $(\mathcal{G}, p)$ is \textit{infinitesimally bearing rigid} if the infinitesimal bearing motions are all trivial.
\end{definition}

\begin{Lemma}
A framework $(\mathcal{G}, p)$ is \textit{infinitesimally bearing rigid} in $\mathbb{R}^d$ if and only if 

(a) $\mathrm{Rank}(R_b(p))=nd-d-1$;

(b) $\mathrm{Null}(R_b(p)) = \mathrm{span}\{ \mathbf{1}_n \otimes I_d, p \}$.
\end{Lemma}
To sum up, any \textit{infinitesimal bearing rigid} framework can be uniquely determined up to trivial motions.

We introduce an orthogonal projection operator \cite{zhao2015bearingtac} that will be useful for later analysis. For any nonzero vector $x \in \mathbb{R}^{d} \ (d \ge 2)$, we define the projection operator $P_x$ as
\begin{equation*}\label{operator}
P_{x}:=I_d - \frac{x}{||x||}\frac{x^T}{||x||},
\end{equation*}
where $P_{x}$ is an orthogonal projection matrix that projects any vector onto the orthogonal complement of $x$. We note that the orthogonal projection matrix $P_{x}$ has the following properties; 1) symmetric, i.e., $P_{x}=P_{x}^T$, 2) positive semidefinite, i.e., $P_x \ge 0$, 3) idempotent, i.e., $P_{x}=P_{x}^2$. Moreover, $\mathrm{Null}(P_{x})=$ span$\{ x \} $ and the eigenvalues of $P_{x}$ are $\{ 0, 1, \ .\ .\ . \ , 1 \}$. The orthogonal projection matrix $P_{x}$ can be utilized to evaluate whether two vectors are parallel or not.
\begin{Lemma}{\cite{zhao2015bearingtac}}\label{projection}
Two nonzero vectors $x, y \in \mathbb{R}^d$ are parallel if and only if $P_{x}y = P_{y}x=0$.
\end{Lemma}

From the bearing rigidity theory, the bearing rigidity matrix can be expressed using the orthogonal projection matrix as 
\begin{equation*}\label{usef}
R_b(p)=diag \left( \frac{P_{g_k}}{||z_k||} \right) \bar{H}.
\end{equation*}
%Using the result of Lemma \ref{jungin} and (\ref{usef}), the following lemma shows the rank condition of the rigidity matrix of infinitesimal bearing rigid framework.

%\begin{lemma}\label{yeri}
%A framework $(\mathcal{G}, p)$ is infinitesimally bearing rigid if and only if $\rank(\tilde{R}_b(p))=\rank(\tilde{R}_b(p)\tilde{R}_b(p)^T)=\rank({R}_b(p)R_b(p)^T)=nd-d-1$.
%\end{lemma}
We next introduce the notion of bearing localizability that can determine the unique formation configuration. 
%The bearing localizability will be useful for analysis of the leader-fixed follower formation control (Section \ref{sec33}) and network localization (Section \ref{sec4}) systems.
\begin{definition}{\cite{zhao2016localizability}}\label{localize}
A framework $(\mathcal{G}, p)$
%\footnote{Note that the $(\mathcal{G}, p^{*})$ means the target formation (respectively, stationary localization network) for the leader-follower formation control (respectively, network localization) system.}
is \textit{bearing localizable} if the configuration $p$ is uniquely determined by the given relative bearing $\{ g_{ij} \} _{(i, j) \in \mathcal{E}}$ and the given stationary positions of the leader agents $\{p_i\}_{i \in \mathcal{V}_L}$. 
\end{definition}
%We note that the $(\mathcal{G}, p^{*})$ means the target formation (respectively, stationary localization network) for the leader-follower formation (respectively, network localization) system.

From \cite{zhao2016localizability}, the bearing Laplacian matrix $\mathcal{B}$ for $(\mathcal{G}, p)$ is defined as
\begin{equation*}\label{sungha}
[\mathcal{B}(\mathcal{G}(p))]_{ij} = \begin{cases} \mathbf{0}_{d \times d}, & i \neq j, (i, j) \notin \mathcal{E} \\ -P_{g_{ij}},   & i \neq j, (i, j) \in \mathcal{E}  \\ \sum_{k \in \mathcal{N}_i}P_{g_{ik}}, & i = j, i \in \mathcal{V}.  \end{cases}
\end{equation*}
Also, the bearing Laplacian matrix $\mathcal{B}$ for the leader-fixed follower formation system can be partitioned as
\begin{equation*}\label{mmy}
\mathcal{B} = \begin{bmatrix} \mathcal{B}_{ll} & \mathcal{B}_{lf} \\ \mathcal{B}_{fl}  & \mathcal{B}_{ff}  \end{bmatrix} \in \mathbb{R}^{dn \times dn},
\end{equation*}
where $\mathcal{B}_{ll} \in \mathbb{R}^{dn_l \times dn_l}, \mathcal{B}_{lf} = \mathcal{B}_{fl}^T \in \mathbb{R}^{dn_l \times dn_f}, \mathcal{B}_{ff} \in \mathbb{R}^{dn_f \times dn_f}$. Also, the bearing Laplacian matrix $\mathcal{B}$ can be represented as $\mathcal{B}=\bar{H}^Tdiag(P_{g_k})\bar{H}$.

The following lemma states the mathematical condition of the bearing Laplacian matrix that characterizes the bearing localizability of the framework.
\begin{Lemma}{\cite{zhao2016localizability}}\label{dahan}
For the leader-fixed follower formation system, a framework $(\mathcal{G}, p)$ is \textit{bearing localizable} if and only if $\mathcal{B}_{ff}$ is positive definite.
\end{Lemma}

For later stability analysis, we introduce the following useful lemmas.
%\begin{definition}{\cite{zhao2016bearing}}
%The centroid of the formation system is defined as 
%\begin{equation}\label{centroid}
%\bar{p}:=\frac{1}{n} \sum_{i=1}^n p_i.
%\end{equation}
%Note that the centroid $\bar{p}$ also can be written as $\bar{p} = (\mathbf{1}_n \otimes I_d)^Tp/n$, where $\mathbf{1}_n = [1,\ .\ .\ . \ , 1]^T \in \mathbb{R}^{n}$. 
%The scale of the formation system is defined as
%\begin{equation}\label{sscale}
%\mathbf{s}:=\sqrt{\frac{1}{n}\sum_{i=1}^n ||p_i - \bar{p}||^2}.
%\end{equation}
%Note that the scale $\mathbf{s}$ also can be written as $\mathbf{s} = ||p - \mathbf{1}_n\otimes \bar{p}||/\sqrt{n}$.
%\end{definition}
\begin{Lemma}{\cite{harville1998matrix}}\label{migu}
For a symmetric and positive semidefinite matrix $P$, we denote the smallest positive eigenvalue of $P$ as $\lambda_{\min}^{+}(P)$ and the orthogonal complement of $\mathrm{Null}(P)$ as $\mathrm{Null}(P)^{\perp}$. Then, for any non-zero vector $x \in \mathrm{Null}(P)^{\perp}$, we have $0<\lambda_{\min}^{+}(P)||x||^2 \le x^TPx$.
\end{Lemma}
\begin{Lemma}\cite{carless1981continuous}\label{haeb}
For a system $\dot{x}=f(x)$, where $f$ is locally Lipschitz in $x$, assume that there exists a continuously differentiable function $V(x)$ such that 
\begin{equation*}
\begin{split}
\alpha_1(||x||)& \le V(x)\le \alpha_2(||x||),\\
\dot{V}(x) & \le -\alpha_3(||x||)+\beta,
\end{split}
\end{equation*}
where $\beta$ is a positive constant, $\alpha_1$ and $\alpha_2$ are class $\mathbb{K}_{\infty}$ functions, and $\alpha_3$ is a class $\mathbb{K}$ function. Then, the solution $x$ of $\dot{x}=f(x)$ is ultimately bounded.
\end{Lemma}
\begin{Lemma}{\cite{bernstein2005matrix}}\label{young}
For any non-zero vectors $x, y$ and positive real numbers $p, q, \epsilon$ such that $1/p + 1/q=1$, the following inequality holds
\begin{equation*}
x^Ty \le \frac{\epsilon^p}{p}||x||^p + \frac{\epsilon^{-q}}{q}||y||^q.
\end{equation*}
%If $a$ and $b$ are non-negative real numbers and $p$ and $q$ are positive real numbers such that $1/p + 1/q=1$, then $ab \leq a^p/p+b^q/q$.
\end{Lemma}
%%%%%%%%%%%%%%%%%%%%%%%%%%%%%%%%%%%%%%%%%%%%%
\section{Bearing-based formation control with exogenous disturbances}\label{sec3}
This section studies the bearing-based formation shape stabilization problem for two formation control systems in the presence of the exogenous disturbances. We first analyze the stability of the leaderless bearing-based formation control system in Section \ref{sec32}. Afterward, we analyze the stability of the leader-fixed follower bearing-only formation control system in Section \ref{sec33}. The analysis for the leader-fixed follower formation control system is based on our earlier work in \cite{bae2019leader}. However, the scale of the formation was assumed to be upper-bounded in \cite{bae2019leader}. We further extend our previous results by deleting such an assumption in this paper. 

\subsection{Problem statement}
%In this section, we formulate the bearing-based formation stabilization problem in the presence of the exogenous disturbance. 
The agent dynamics is assumed to be modeled by a single-integrator model in the presence of the exogenous disturbances
\begin{equation}\label{single}
\dot{p}_i = u_i + f_i,
\end{equation}
where $\dot{p}_i, u_i, f_i$ denote the velocity, the control input and the time-varying exogenous disturbance of agent $i$. We assume that each agent can sense a common global reference frame. In real-world applications, a common global reference frame of the agent can be measured by on-board sensors, e.g., compass. Also, we assume that each agent has bearing sensing capability that measures the relative bearing information of neighboring agents. For example, the measured relative bearing vector of agent $j$ from agent $i$ is denoted by $g_{ij}$. The desired relative bearing vector set to achieve the target formation shape of $(\mathcal{G}, p^*)$ is denoted by $\mathfrak{G}=\{g_{ij}^* \in \mathbb{R}^d \ | \ {\forall(i, j) \in \mathcal{E}} \}$ with $g^{*}_{ij}=-g^{*}_{ji}$. Then, the bearing-based formation shape stabilization problem in this paper is stated as follows:
\begin{Problem}{(\textit{Bearing-based Formation Control}):}\label{prob}
For a given feasible desired relative bearing vector set $\mathfrak{G}$ and the initial formation $\mathcal{G}(p(0))$, design a distributed bearing-based control law $u_i(t)$ such that $g_{ij}(t) \rightarrow g_{ij}^{*}, \forall(i, j) \in \mathcal{E}$.
\end{Problem}

We consider Problem \ref{prob} with the exogenous disturbances in the following subsections. In the real-world, there exist various sources of exogenous disturbances and uncertainties, then the formation system may not achieve the formation control objective, which can be called \textit{bearing formation error}. Therefore, it is worth analyzing the effect of the exogenous disturbances on the \textit{bearing formation error}. We investigate the issues mentioned above by way of analyzing stability and computing the upper-bound of \textit{bearing formation error} in later.
%%%%%%%%%%%%%%%%%%%%%%%%%%%%%%%%%%%%%%%%%%%%%%%%%%%%%%%%%%%%%%%
\subsection{Stability analysis for leaderless bearing-based formation control system}\label{sec32}
%%%%%%%%%%%%%%%%%%%%%%%%%%%%%%%%%%%%%%%%%%%%%
%\subsection{Distributed adaptive gradient controller}
We consider leaderless bearing-based formation control system with the exogenous disturbances. The target formation is defined by the feasible desired relative bearing vector set $\mathfrak{G}$ and assumed to be \textit{infinitesimally bearing rigid}.
%\begin{assumption}\label{cham}
%The target formation is \textit{infinitesimally bearing rigid}.
%\end{assumption}

Inspired by the distributed bearing-only control law \cite{zhao2015bearingtac}, we consider the bearing-based formation control law which uses relative bearing information with auxiliary range measurements in the presence of the exogenous disturbances as
\begin{equation}\label{songho}
\dot{p}_i(t) = u_i(t)+f_i(t) = - \sum_{j \in \mathcal{N}_i}\frac{P_{{g}_{ij}(t)}}{||z_{ij}(t)||}g^{*}_{ij}+f_i(t), i \in \mathcal{V}
\end{equation}
where $f_i(t)  \in \mathbb{R}^{d} $ denotes the time-varying exogenous disturbance vector of agent $i$. 

From (\ref{songho}), the overall dynamics for the leaderless bearing-based formation control system is derived as 
\begin{equation}\label{yuk}
\dot{p}=u+f=\bar{H}^Tdiag \left( \frac{P_{g_k}}{||z_k||} \right)g^* +f = {R}_b^Tg^* +f,
\end{equation}
where $g^*\in \mathbb{R}^{md}$ and $f \in \mathbb{R}^{nd}$ denote the desired relative bearing vector and the exogenous disturbances.

\begin{Assumption}\label{jiso}
The time-varying exogenous disturbances are uniformly upper-bounded, i.e., $||f(t)|| \le \sqrt{{\sum_{i=1}^n}v_i^2} = \mathcal{F}$, where $\mathcal{F}$ is a small positive constant.
\end{Assumption}

Intuitively, for the leaderless bearing-based formation control system (\ref{songho}), the entire formation may have continuous movement in space, e.g., translational motion in two-dimensional space, in the presence of the exogenous disturbances. 
In this aspect, we define the \textit{bearing formation error} $e_a$ as 
\begin{equation}\label{overberr}
e_a := g - g^*,
\end{equation}
 where $e_a =[e_{a_{1}}^T, \ e_{a_{2}}^T,\ .\ .\ . \ ,e_{a_{m}}^T]^T \in \mathbb{R}^{md}$. 
%\begin{remark}
%n the previouis paper~, the error dynamics is $p-p^{*}$~
%\end{remark}.
From the definition of the \textit{bearing formation error}, it follows that the convergence of the \textit{bearing formation error} $e_a$ to zero means that the leaderless bearing-based formation converges to the target formation shape.
%We note that the Euclidean square norm of the \textit{bearing formation error} $e_b$ can be upper-bounded as
%\begin{equation}\label{cheol}
%||e_{b_{k}}||^2 = ||g_k-g_k^*||^2 = 2(1-g_k^{*T}g_k)\le 4.
%\end{equation}
%\begin{equation}\label{cheol2}
%||e_{b}||^2 = ||e_{b_{1}}||^2 + ||e_{b_{2}}||^2 + ... + ||e_{b_{m}}||^2 \le 4m.
%\end{equation}
%From the definition of the \textit{bearing formation error}, it follows that the convergence of the \textit{bearing formation error} $e_b$ to zero means that the leaderless bearing-based formation converges to the target formation. 
\begin{Lemma}\label{dasin}
From (\ref{overberr}), the \textit{bearing formation error} $e_a$ for arbitrary edge $k$ is defined as $e_{a_k} := g_k - g_k^*$. Then, the following equations hold 
\begin{equation*}
1-g_k^{*^T}g_k = \frac{||e_{a_k}||^2}{2},
\end{equation*}
\begin{equation*}
1+g_k^{*^T}g_k = 2-\frac{||e_{a_k}||^2}{2}.
\end{equation*}
\begin{proof}
From $e_{a_k} = g_k - g_k^*$, it follows $||e_{a_k}||^2 = 2(1-g_k^{*T}g_k)$. Therefore, we have $1-g_k^{*^T}g_k = ||e_{a_k}||^2/2$. Accordingly, $1+g_k^{*^T}g_k = 2-  ||e_{a_k}||^2/2$.
\end{proof}
\end{Lemma}

From (\ref{overberr}), the error dynamics for the leaderless bearing-based formation control system is given as
\begin{equation}\label{berrdyn}
\begin{split}
\dot{e}_a&=\frac{\partial{e_a}}{\partial{p}}\dot{p} \\
&=R_b{R}_b^Tg^*+R_{b}f \\ 
&= R_b{R}_b^T(g-e_a)+R_{b}f \\
&= -R_b{R}_b^Te_a+R_{b}f,
\end{split}
\end{equation}
where $R_b{R}_b^Tg = 0$ since we have $g \in \mathrm{Null}({R}_b^T)$. 

%We next introduce a new bearing formation error variable $e_a^N$ for later stability analysis.

%\begin{Lemma}\label{cnblue}
%Let us define a new bearing formation error variable $e_a^{N}:= diag(P_{g_{k}})e_a$. Then, $e_a^{N} =0$ if and only if $g = g^*$ or $g = -g^*$. 
%\begin{proof}
%Since $e_a^{N}= diag(P_{g_{k}})e_a = -diag(P_{g_{k}})g^*$, $e_a^N=0$ if and only if $g = g^*$(\textit{desired}) or $g = -g^*$(\textit{undesired}).
%\end{proof}
%\end{Lemma}
%In the following stability analysis, an exponential stability of the origin of the error dynamics $e_a$ will be proved by showing an exponential stability of the origin of $e_a^N$. As noted in Lemma \ref{cnblue}, $e_a^N=0$ implies that there are two isolated equilibrium states which are \textit{desired one}, i.e., $g = g^*$, and \textit{undesired one}, i.e., $g = -g^*$. To avoid the undesired equilibrium state, we confine our results to local stability analysis only.

Also, we state the following assumption for later stability analysis.
\begin{Assumption}\label{suff}
The positions of the neighboring agents are not collocated during formation evolvement.
\end{Assumption}

The following Lyapunov stability analysis shows the local asymptotic bound of the \textit{bearing formation error} $e_a$.

\begin{Theorem}
Consider the leaderless bearing-based formation control system (\ref{songho}) with the exogenous disturbances. For infinitesimally bearing rigid formation  under Assumptions \ref{jiso}-\ref{suff}, the \textit{bearing formation error} $e_a$ is locally asymptotically convergent and bounded for sufficiently small exogenous disturbances satisfying $\mathcal{F}\le \sqrt{{\lambda_{\min_t}^{+}(R_b{R}_b^T)^2}/{\lambda_{\max_t}(R_b{R}_b^T)}}$. Furthermore, the \textit{bearing formation error} $e_a$ locally asymptotically converges to the following bound set $\mathcal{S}_a$
\begin{equation}\label{na}
\mathcal{S}_a=\left\{ e_a:||e_a||^2 \le 2-2\sqrt{1- \frac{{\lambda_{\max_t}(R_b{R}_b^T)}\mathcal{F}^2}{{\lambda_{\min_t}^{+}(R_b{R}_b^T)^2}}} \right\},
\end{equation}
where $\lambda_{\min_{t}}^{+}(R_b{R}_b^T)$ and $\lambda_{\max_t}(R_bR_b^T)$ are the smallest positive and maximum eigenvalue of $R_bR_b^T$ for all time, which are positive constants.
%where $\varepsilon$ is a positive constant satisfying $0<\varepsilon<1$.
\begin{proof}
We consider the following Lyapunov candidate function
\begin{equation}\label{yhwa}
V_a = \frac{1}{2}||e_a||^2, 
\end{equation}
for $e_a(0) \in \phi$, where $\phi = \{e_a : ||e_a||^2 \le \varphi \}$ for small $\varphi$, such that the initial formation is close to the target formation shape in the set $\phi$.  
From (\ref{yhwa}), the time derivative of $V_a$ is derived as
\begin{equation}\label{hayong}
\begin{split}
\dot{V}_a  &= e_a^T\dot{e}_a\\
%&= -g^{*T}[R_bR_b^Tg^{*} + R_bf] \\
&= -e_a^TR_b{R}_b^Te_a + e_a^TR_bf\\
&=-e_a^Tdiag\left( \frac{P_{g_k}}{||z_k||} \right) \bar{H}\bar{H}^Tdiag\left( \frac{P_{g_k}}{||z_k||} \right)e_a + e_a^TR_bf \\
&= -e_a^{N^T}diag\left( \frac{P_{g_k}}{||z_k||} \right) \bar{H}\bar{H}^Tdiag\left( \frac{P_{g_k}}{||z_k||} \right)e_a^{N} + e_a^{N^T}R_bf  \\
&= -e_a^{N^T}R_b{R}_b^Te_a^{N} + e_a^{N^T}R_bf,
\end{split}
\end{equation}
where we define $e_a^N := diag(P_{g_k})e_a$, and have the idempotent property for $diag({P_{g_k}})$, i.e., $diag({P_{g_k}}) = diag({P_{g_k}})^2$. 
From $e_a^N = diag({P_{g_k}})e_a$, it follows that $e_a^{N} \in \mathrm{Range}(diag({P_{g_k}}))$ and $e_a^{N} \perp \mathrm{Null}(diag({P_{g_k}}))$.
This fact reflects that $e_a^{N^\perp} \in \mathrm{Null}(diag({P_{g_k}}))$, where $e_a^{N^\perp}$ is the orthogonal complement of $e_a^{N}$. Then, $e_a^{N^\perp} \in \mathrm{Null}(R_b{R}_b^T)$ since we have $\mathrm{Null}(diag({P_{g_k}}))$ $\subseteq \mathrm{Null}(R_b{R}_b^T)$. Finally, it follows that $e_a^{N} \in \mathrm{Null}(R_b{R}_b^T)^{\perp}$. Therefore, based on Lemma \ref{migu}, (\ref{hayong}) can be written as
\begin{equation}\label{soo}
\dot{V}_a \le  -{\lambda_{\min}^{+}(R_b{R}_b^T})||e_a^{N}||^2 + e_a^{N^T}R_bf.
\end{equation}
Based on Lemma \ref{young}, which is young's inequality, (\ref{soo}) can be written as
\begin{equation}\label{sunnn}
\begin{split}
\dot{V}_a&\le -{\lambda_{\min}^{+}(R_b{R}_b^T})||e_a^{N}||^2 +  \frac{e_a^{N^T}R_bR_b^Te_a^{N}}{\gamma^{2}} + \frac{\gamma^2}{4}||f||^2 \\
&\le -{\lambda_{\min}^{+}(R_b{R}_b^T})||e_a^{N}||^2 +  \frac{\lambda_{\max}(R_bR_b^T)}{\gamma^{2}}||e_a^{N}||^2 + \frac{\gamma^2\mathcal{F}^2}{4}\\
&\le \left( \frac{\lambda_{\max_t}(R_bR_b^T)}{\gamma^2} - {\lambda_{\min_{t}}^{+}(R_b{R}_b^T)} \right)||e_a^{N}||^2+ \frac{\gamma^2\mathcal{F}^2}{4},
%&= -\frac{\lambda_{\min}^{+}(R_b\tilde{R}_b^T+\tilde{R}_bR_b^T)}{2}e^TP_ge+e^TR_{b}f \\
%&= -\frac{\lambda_{\min}^{+}(R_b\tilde{R}_b^T+\tilde{R}_bR_b^T)}{2}g^{{*}T}P_gg^{*}+e^TR_{b}f,
\end{split}
\end{equation}
where $\lambda_{\min_{t}}^{+}(R_b{R}_b^T)$ and $\lambda_{\max_t}(R_bR_b^T)$ are the smallest positive and maximum eigenvalue of $R_bR_b^T$ for all time, which are positive constants. Also, $||f||$ is bounded as $\mathcal{F}$ under Assumption \ref{jiso}. 
We define the constant $\gamma$ in (\ref{sunnn}) as
\begin{equation}\label{myung}
\gamma^{-2}:=\frac{{\lambda_{\min_t}^{+}(R_b{R}_b^T)}}{2\lambda_{\max_t}(R_bR_b^T)}>0.
\end{equation}
Then, (\ref{sunnn}) can be written as
\begin{equation}\label{dddd}
\dot{V}_a\le -\frac{{\lambda_{\min_t}^{+}(R_b{R}_b^T)}}{2}||e_a^{N}||^2 + \frac{{\lambda_{\max_t}(R_b{R}_b^T)}\mathcal{F}^2}{2{\lambda_{\min_t}^{+}(R_b{R}_b^T)}}.
\end{equation}
%%%%%%%%%%%%%%%%%%%%
From $e_a^N = diag({P_{g_k}})e_a$, $||e_a^N||^2$ can be expressed in terms of $||e_{a}||^2$ as follows:
\begin{equation}\label{jisub}
\begin{split}
||e_a^N||^2 &= e_a^Tdiag({P_{g_k}})e_a  =  g^{{*}^T}diag({P_{g_k}})g^{*} \\
&= \sum_{k=1}^m{g_k^{*^T}P_{g_k}g_k^{*}} \\
&= \sum_{k=1}^m{1-(g_k^{*^T}g_k)^2} \\
&= \sum_{k=1}^m{(1 + g_k^{*^T}g_k)(1 - g_k^{*^T}g_k)} \\
&= \sum_{k=1}^m{||e_{a_k}||^2 - \frac{||e_{a_k}||^4}{4}}\ge ||e_a||^2 - \frac{||e_a||^4}{4}, \\
\end{split}
\end{equation}
where we have used the result of Lemma \ref{dasin} to get the last equation.
Then, (\ref{dddd}) can be written as
\begin{equation}\label{junki}
\dot{V}_a\le -\frac{{\lambda_{\min_t}^{+}(R_b{R}_b^T)}}{2} \left( ||e_a||^2 - \frac{||e_a||^4}{4} \right) + \frac{{\lambda_{\max_t}(R_b{R}_b^T)}\mathcal{F}^2}{2{\lambda_{\min_t}^{+}(R_b{R}_b^T)}}.
\end{equation} 
Since (\ref{junki}) is a quadratic inequality in terms of $||e_a||^2$, it follows that $\dot{V}_a<0$ if
\begin{equation}\label{xii}
 2-2\sqrt{1- \psi}<||e_a||^2<2+2\sqrt{1- \psi},
\end{equation}
where we denote $\psi = {\lambda_{\max_t}(R_b{R}_b^T)}\mathcal{F}^2/{\lambda_{\min_t}^{+}(R_b{R}_b^T)^2}$.
Therefore, from (\ref{xii}), assuming that the initial formation starts in the set $\phi$, i.e., $e_a(0) \in \phi$, where $\phi = \{e_a : ||e_a||^2 \le \varphi \}$, such that
\begin{equation*}
\varphi < 2+2\sqrt{1- \frac{{\lambda_{\max_t}(R_b{R}_b^T)}\mathcal{F}^2}{{\lambda_{\min_t}^{+}(R_b{R}_b^T)^2}}}.
\end{equation*}
We have, $\dot{V}_a<0$ if
\begin{equation}
||e_a||^2 > 2-2\sqrt{1- \frac{{\lambda_{\max_t}(R_b{R}_b^T)}\mathcal{F}^2}{{\lambda_{\min_t}^{+}(R_b{R}_b^T)^2}}}.
\end{equation}
As a consequence, it follows that the \textit{bearing formation error} $e_a$ is locally asymptotically convergent and bounded to the set $\mathcal{S}_a$ in (\ref{na}).
\end{proof}
\end{Theorem}

%\begin{Remark}
%By choosing $\varepsilon =  1/2$, the upper-bound set $\mathcal{S}_a$ in (\ref{na}) becomes a minimum as follows:
%\begin{equation}
%\mathcal{S}_a^{\min}=\left\{ e_a:{\sum_{k=1}^{m}||e_{a_k}||^2-\frac{||e_{a_k}||^4}{4}}\le \frac{4{\lambda_{\max}(R_b{R}_b^T)}\mathcal{F}^2}{{\lambda_{\min}^{+}(R_b{R}_b^T)^2}} \right\}.
%\end{equation}
%\end{Remark}

Note that the upper-bound set $\mathcal{S}_a$ in (\ref{na}) consists of the state-dependent matrix $RR^T$. Therefore, a precise bound is incomputable; however, some guide can be provided for a system design by way of computing $\mathcal{S}_a$, e.g., optimal graph design is required to minimize $\lambda_{\max_t}(R_b{R}_b^T)$ or maximize $\lambda_{\min_t}^{+}(R_b{R}_b^T)$ for smaller upper-bound of the \textit{bearing formation error} $e_a$. 
%In this aspect, computable upper-bound set of \textit{bearing formation error} is expected by considering fixed-leader agents since the error dynamics becomes concise.

%\newpage
%%%%%%%%%%%%%%%%%%%%%%%%%%%%%%%%%%%%%%%%%%%%%%%%%%%%%%%%%%%%%%
\subsection{Stability analysis for leader-fixed follower bearing-only formation control system}\label{sec33}
We consider Problem \ref{prob} for the leader-fixed follower bearing-only formation control system with the exogenous disturbances. %We note that the analysis on this section is based on \ci te{}%We note that the leader agents are stationary for all time, i.e., $p_i(t) = p_i^*, i \in \mathcal{V}_L$. 
Unlike the previous subsection, a precise bound of \textit{bearing formation error} is expected to be computed by considering fixed-leader agents since Lyapunov stability analysis is governed by the bearing Laplacian matrix, which is a constant matrix, as will be shown later. For the leader-fixed follower formation control system, the position vector set can be divided into $p= [p_{l}^T,  p_{f}^T]^T \in \mathbb{R}^{nd}$, where $p_l$ and $p_f$ denote the position vectors of the leader and follower agents, respectively. Note that $p_l(t) = p_l^{*}$, which means that the leader agents are stationary to their desired positions during formation evolvement.
Then, the \textit{bearing localizable} target formation is defined by the feasible desired relative bearing vector set $\mathfrak{G}$ and the stationary positions of the leader agents $p_l^*$. 
%\begin{assumption}\label{yang}
%The target formation is \textit{bearing localizable} and consequently, at least two leader agents exist, i.e., $\mathcal{B}_{ff}$ in (\ref{mmy}) is positive definite.
%\end{assumption}
From \cite{zhao2016localizability}, note that at least two leader agents need to exist to ensure the bearing localizability.

The formation shape stabilization problem (Problem \ref{prob}) is converted to the formation configuration stabilization problem for the leader-fixed follower  formation control system since the configuration of the \textit{bearing localizable} target formation is uniquely determined in space, as stated in Definition \ref{localize}. Even if we consider the exogenous disturbances to the formation control system, the entire formation is not moving in space since leader agents keep their stationary positions. Therefore, convergence of the follower agents to their desired positions, i.e., $p_i(t) \rightarrow p_i^*$ for $i \in \mathcal{V}_F$, means that the formation system converges to the desired configuration $p^*$. In this aspect, we define the \textit{bearing formation error} as 
\begin{equation}\label{error}
e_b(t) := p(t) - p^*,
\end{equation}
where $e_b =[e_{b_1}^T, \ e_{b_2}^T,\ .\ .\ . \ ,e_{b_n}^T]^T = [0, e_{b_f}^T]^T \in \mathbb{R}^{nd}$ since $p_l(t) = p_l^*$, and $p^*$ is the desired configuration of the target formation $(\mathcal{G}, p^*)$. From (\ref{error}), the error dynamics is given as
\begin{equation}
\dot{e}_b = \dot{p}.
\end{equation}
%The following definition introduces the centroid and the scale of the formation system.
%\begin{definition}[\cite{zhao2016bearing}]
%The centroid of the formation system is defined as 
%\begin{equation}\label{centroid}
%\bar{p}:=\frac{1}{n} \sum_{i=1}^n p_i.
%\end{equation}
%We note that the centroid $\bar{p}$ also can be written as $\bar{p} = (\mathbf{1} \otimes I_d)^Tp/n$, where $\mathbf{1} = [1,\ .\ .\ . \ , 1]^T \in \mathbb{R}^{n}$. 
%The scale of the formation system is defined as
%\begin{equation}\label{scale}
%\mathbf{s}:=\sqrt{\frac{1}{n}\sum_{i=1}^n ||p_i - \bar{p}||^2}.
%\end{equation}
%We note that the scale $\mathbf{s}$ also can be written as $\mathbf{s} = ||p - \mathbf{1}\otimes \bar{p}||/\sqrt{n}$.
%\end{definition}

Based on the distributed bearing-only control law that ensures global convergence \cite{zhao2018revisit, tron2016bearing, zhao2019bearing}, we consider the leader-fixed follower bearing-only formation control system with the exogenous disturbances as
\begin{equation}\label{songgho}
\begin{split}
\dot{p}_{i}(t) &= 0, i \in \mathcal{V}_L \\
\dot{p}_i(t) & = \sum_{j \in \mathcal{N}_i}(g_{ij}(t) - g^{*}_{ij})+f_i(t),  i \in \mathcal{V}_F
\end{split}
\end{equation}
where $f_i(t) \in \mathbb{R}^{d}$ denotes the time-varying exogenous disturbance vector of agent $i$. %In addition, the exogenous disturbance satisfies the following assumption.
%\begin{assumption}\label{jisoo}
%The exogenous disturbance is upper-bounded, i.e., $||f_i||\le v_i, i \in \mathcal{V}_F$, where $v_i$ is a positive constant. 
%\end{assumption}

From (\ref{songgho}), the overall dynamics for the leader-fixed follower bearing-only formation control system is derived as 
\begin{equation}\label{yukk}
\dot{p}=u+f=-\begin{bmatrix} 0 & 0 \\ 0 & I_{dn_f} \end{bmatrix}\bar{H}^T(g - g^*) + f,
\end{equation}
where $g^*\in \mathbb{R}^{md}$, $f \in \mathbb{R}^{nd}$ denote the desired relative bearing vector and the exogenous disturbance vector.
\begin{Assumption}\label{jisoo}
The time-varying exogenous disturbances are uniformly upper-bounded, i.e., $||f|| \le \sqrt{{\sum_{i=n_l + 1}^n}v_i^2} = \mathcal{F}$, where $\mathcal{F}$ is a small positive constant.
\end{Assumption}

We state the following lemmas for later stability analysis.

\begin{Lemma}{\cite{zhao2018revisit}}\label{saam}
Under Assumption \ref{suff}, it holds 
\begin{equation}\label{cd}
(p^{*})^T\bar{H}^T(g-g^*)\leq0,
\end{equation}
\begin{equation}\label{dc}
(p-p^{*})^T\bar{H}^T(g-g^*)\geq0,
\end{equation}
where the equalities hold if and only if $g = g^*$.
\end{Lemma}

\begin{Lemma}{\cite{zhao2018revisit}}\label{saamsu}
Under Assumption \ref{suff}, it holds 
\begin{equation}\label{bor}
p^T\bar{H}^T(g-g^*) \ge \frac{1}{2\max_{k}||z_k||}p^T\mathcal{B}p,
\end{equation}
where $\mathcal{B}$ is the bearing Laplacian matrix for $(\mathcal{G}, p^*)$.
\end{Lemma}

The following Lyapunov stability analysis shows the global asymptotic bound of the \textit{bearing formation error} $e_b$.
\begin{Theorem}\label{bbmain}
Consider the leader-fixed follower bearing-only formation control system (\ref{songgho}) with the exogenous disturbances. For bearing localizable formation system under Assumptions \ref{suff}-\ref{jisoo}, the \textit{bearing formation error} $e_b$ is globally asymptotically convergent and bounded for the exogenous disturbances satisfying $\mathcal{F} <{\sqrt{\varepsilon(\lambda_{\min}(\mathcal{B}_{ff})-\varepsilon)}}/{||\bar{H}||}$. Furthermore, the \textit{bearing formation error} $e_b$ globally asymptotically converges to the following bound set $\mathcal{S}_b$
\begin{equation}\label{nna}
\mathcal{S}_b= \left\{ e_b : ||e_b|| \le \frac{||p^*||||\bar{H}||\mathcal{F}}{\sqrt{\varepsilon(\lambda_{\min}(\mathcal{B}_{ff})-\varepsilon)} - ||\bar{H}||\mathcal{F}} \right\},
\end{equation}
where $\varepsilon$ is a positive constant satisfying $0<\varepsilon<\lambda_{\min}(\mathcal{B}_{ff})$. 
%where $\delta$ is a positive constant.
\begin{proof}
We consider the following Lyapunov candidate function
\begin{equation}\label{cull}
V_{{b}}  =  \frac{1}{2}||e_b||^2.
\end{equation}
From (\ref{error}) and (\ref{yukk}), the time derivative of $V_b$ is derived as
\begin{equation}\label{hayoong}
\begin{split}
\dot{V}_{{b}} &=  e_b^T\dot{e_b} \\
&= e_b^T\dot{p} \\
&=-e_b^T{\begin{bmatrix} 0 & 0 \\ 0 & I_{dn_f} \end{bmatrix}\bar{H}^T(g - g^*) + e_b^Tf} \\
&= -e_b^T\bar{H}^T(g - g^*) + e_b^Tf,
\end{split}
\end{equation}
where $e_b = [0, e_{bf}^T]^T$ since $p_l(t) = p_l^*$. From (\ref{cd}) in Lemma \ref{saam}, (\ref{hayoong}) can be written as
\begin{equation}\label{gu}
\begin{split}
\dot{V}_{{b}} &= -(p - p^*)^T\bar{H}^T(g - g^*) + e_b^Tf \\
&\leq -p^T\bar{H}^T(g - g^*) + e_b^Tf.
\end{split}
\end{equation}
From (\ref{bor}) in Lemma \ref{saamsu}, (\ref{gu}) can be written as
\begin{equation}\label{ikon}
\begin{split}
\dot{V}_{{b}} &\le - \frac{1}{2\max_{k}||z_k||}p^T\mathcal{B}p + e_b^Tf, \\
%&\leq - \frac{1}{4n\bar{\mathbf{s}}}p^T\mathcal{B}p + e_a^Tf,
\end{split}
\end{equation}
where $\mathcal{B}$ is the bearing Laplacian matrix for the target formation $(\mathcal{G}, p^*)$. Note that $p^T\mathcal{B}p = (p - p^{*})^T\mathcal{B}(p - p^{*}) = e_b^T\mathcal{B}e_b$ since $\mathcal{B}p^*=0$. Therefore, (\ref{ikon}) can be written as
\begin{equation}\label{don}
\begin{split}
\dot{V}_b &\le - \frac{1}{2\max_{k}||z_k||}e_b^T\mathcal{B}e_b + e_b^Tf \\
&= - \frac{1}{2\max_{k}||z_k||}e_{bf}^T\mathcal{B}_{ff}e_{bf} + e_b^Tf \\
&\le - \frac{\lambda_{\min}(\mathcal{B}_{ff})}{2\max_{k}||z_k||}||e_b||^2 + e_b^Tf, \\
%&\le- \frac{\lambda_{\min}(\mathcal{B}_{ff})}{2\max_{k}||z_k||}||e_a||^2 + \frac{\gamma^{-2}}{4}||e_a||^2 + \gamma^2 ||f||^2, \\
%&= -(\frac{\lambda_{\min}(\mathcal{B}_{ff})}{4n\bar{\mathbf{s}}} - \frac{\gamma^{-2}}{4})||e_a||^2 + \gamma^2||f||^2,
\end{split}
\end{equation}
where $\lambda_{\min}(\mathcal{B}_{ff})$ is the smallest eigenvalue of $\mathcal{B}_{ff}$, which is positive from Lemma \ref{dahan}. Note that 
\begin{equation}\label{anji}
\begin{split}
\max_{k}||z_k|| \le ||z|| = ||\bar{H}p|| &= ||\bar{H}(p-p^* + p^*)|| \\
&= ||\bar{H}(e_b + p^*)|| \\
&\le ||\bar{H}||(||e_b|| + ||p^*||).
\end{split}
\end{equation}
%We assume that there exists a finite upper-boundedness $\psi$ such that
%\begin{equation}\label{hyeji}
%||e_a(t)|| \le \psi.
%\end{equation}
%Then, (\ref{anji}) can be written as
%\begin{equation}\label{anjiy}
%\max_{k}||z_k|| \le ||\bar{H}||(\psi + ||p^*||).
%\end{equation}
Based on (\ref{anji}), (\ref{don}) can be written as
\begin{equation}\label{anjih}
\begin{split}
\dot{V}_b &\le- \frac{\lambda_{\min}(\mathcal{B}_{ff})}{2||\bar{H}||(||e_b|| + ||p^*||)}||e_b||^2 +e_b^Tf \\
&\le- \frac{\lambda_{\min}(\mathcal{B}_{ff})}{2||\bar{H}||(||e_b|| + ||p^*||)}||e_b||^2 + \frac{\gamma^{-2}}{4}||e_b||^2 + \gamma^2 ||f||^2 \\
&\le -\left( \frac{\lambda_{\min}(\mathcal{B}_{ff})}{2||\bar{H}||(||e_b|| + ||p^*||)} - \frac{\gamma^{-2}}{4} \right) ||e_b||^2 + \gamma^2 ||f||^2 \\
&= -\left( \frac{\varepsilon}{2||\bar{H}||(||e_b|| + ||p^*||)} \right) ||e_b||^2 + \gamma^2||f||^2,
%&= -(\frac{\lambda_{\min}(\mathcal{B}_{ff})}{4n\bar{\mathbf{s}}} - \frac{\gamma^{-2}}{4})||e_a||^2 + \gamma^2||f||^2,
\end{split}
\end{equation}
where we decompose $e_b^Tf$ based on Lemma \ref{young}. The scalar function $\gamma$ in (\ref{anjih}) is defined as
\begin{equation}\label{sony}
\gamma(||e_b||) := \sqrt{\frac{||\bar{H}||(||e_b|| + ||p^*||)}{2(\lambda_{\min}(\mathcal{B}_{ff}) - \varepsilon)}},
\end{equation}
where $\varepsilon$ is a positive constant satisfying $0<\varepsilon<\lambda_{\min}(\mathcal{B}_{ff})$.
%From the last inequality in (\ref{anjih}), note that the negative value of $-\left( {\varepsilon}/{2||\bar{H}||(||e_a|| + ||p^*||)} \right) ||e_a||^2$ is guaranteed. Then, we note that the \textit{bearing formation error} $e_a$ is ultimately bounded from Lemma \ref{haeb}. 
Then, (\ref{anjih}) can be written as 
\begin{equation}\label{pro}
\dot{V}_b \le -\left( \frac{\varepsilon}{2||\bar{H}||(||e_b|| + ||p^*||)} \right) ||e_b||^2 + \gamma^2\mathcal{F}^2,
\end{equation}
where $||f||$ is upper-bounded as $\mathcal{F}$ under Assumption \ref{jisoo}.
From (\ref{sony}) and (\ref{pro}), it follows that $\dot{V}_b<0$ if
\begin{equation}\label{hyunju}
\begin{split}
||e_b||^2 > \frac{||\bar{H}||^2\mathcal{F}^2(||e_b|| + ||p^*||)^2}{\varepsilon(\lambda_{\min}(\mathcal{B}_{ff})-\varepsilon)} = \psi^2(||e_b|| + ||p^*||)^2,
\end{split}
\end{equation}
where we define $\psi := ||\bar{H}||\mathcal{F}/\sqrt{\varepsilon(\lambda_{\min}(\mathcal{B}_{ff})-\varepsilon)}$, which is a positive constant. Since (\ref{hyunju}) is a quadratic inequality of $||e_b||$, it follows that $\dot{V}_b<0$ if
\begin{equation}\label{seok}
\begin{split}
||e_b|| &> \frac{\psi}{1-\psi}||p^*|| = \frac{||p^*||||\bar{H}||\mathcal{F}}{\sqrt{\varepsilon(\lambda_{\min}(\mathcal{B}_{ff})-\varepsilon)} - ||\bar{H}||\mathcal{F}}.
\end{split}
\end{equation}
Consequently, the \textit{bearing formation error} $e_b$ is globally asymptotically convergent and bounded to the following bound set 
\begin{equation}\label{qed}
||e_b|| \le \frac{||p^*||||\bar{H}||\mathcal{F}}{\sqrt{\varepsilon(\lambda_{\min}(\mathcal{B}_{ff})-\varepsilon)} - ||\bar{H}||\mathcal{F}},
\end{equation}
for small exogenous disturbances satisfying
\begin{equation}\label{exo}
\mathcal{F} < \frac{\sqrt{\varepsilon(\lambda_{\min}(\mathcal{B}_{ff})-\varepsilon)}}{||\bar{H}||}.
\end{equation}
\end{proof}
\end{Theorem}
%We next choose $\varepsilon$ to get the smallest boundedness set of $e_a$ in the following result.
\begin{Corollary}\label{saint} 
By choosing $\varepsilon =  \lambda_{\min}(\mathcal{B}_{ff})/2$, the following results hold 

(a) Smallest bound set of $e_b$
\begin{equation}\label{sam}
\mathcal{S}_b^{\min}= \left\{ e_b : ||e_b|| \le \frac{2||p^*||||\bar{H}||\mathcal{F}}{\lambda_{\min}(\mathcal{B}_{ff}) - 2||\bar{H}||\mathcal{F}} \right\}.
\end{equation}

(b) Maximum upper-bound of the exogenous disturbances.

\begin{proof}
(a) From the bound set in (\ref{nna}), it is obvious that $\sqrt{\varepsilon(\lambda_{\min}(\mathcal{B}_{ff})-\varepsilon)}$ should be maximized to get the smallest bound set of $e_b$. Since $\varepsilon(\lambda_{\min}(\mathcal{B}_{ff})-\varepsilon)$ is a form of quadratic function of $\varepsilon$, $\varepsilon$ should be chosen as $\varepsilon =  \lambda_{\min}(\mathcal{B}_{ff})/2$ to get the smallest bound set of $e_b$, i.e., $\mathcal{S}_b^{\min}$, which becomes (\ref{sam}).

(b) Since $\sqrt{\varepsilon(\lambda_{\min}(\mathcal{B}_{ff})-\varepsilon)}$ is maximized by choosing $\varepsilon = \lambda_{\min}(\mathcal{B}_{ff})/2$, the upper-bound of the exogenous disturbances in (\ref{exo}) is maximized. Therefore, we can handle a larger scale of exogenous disturbances in analysis by choosing $\varepsilon = \lambda_{\min}(\mathcal{B}_{ff})/2$. 
\end{proof}
\end{Corollary}
Even though the leader-fixed follower formation control system is restrictive in system perspective due to the existence of the fixed-leader agents, it is useful for a system design, as stated in the following remarks.
\begin{Remark}\label{yeun}
Theorem \ref{bbmain} showed that the precise bound set $\mathcal{S}_b$ is computable as in (\ref{nna}) with the exogenous disturbances. We note that the set $\mathcal{S}_b$ can be interpreted as the worst case of the \textit{bearing formation error} $e_b$ of the formation system. Therefore, we expect that the set $\mathcal{S}_b$ can be used effectively in real-world applications. For instance, assuming that the system error tolerance is known, we can check the \textit{bearing formation error} is allowable or not for multi-agent formation control tasks by computing the upper-bound set $\mathcal{S}_b$.
\end{Remark}

\begin{Remark}\label{coro}
From (\ref{nna}), it follows that the upper-bound set $\mathcal{S}_b$ is dependent on four system parameters:

(a) The target formation configuration, $||p^*||$,

(b) The graph topology of formation system, $||\bar{H}||$,

(c) The upper-bound of exogenous disturbances, $\mathcal{F}$,

(d) The smallest eigenvalue of $\mathcal{B}_{ff}$, $\lambda_{\min}(\mathcal{B}_{ff})$.

The upper-bound set $\mathcal{S}_b$ becomes smaller by 1) design of graph topology to minimize $||p^*||$ and $||\bar{H}||$, 2) for smaller upper-bound of the exogenous disturbances $\mathcal{F}$, and 3) design to maximize $\lambda_{\min}(\mathcal{B}_{ff})$. Referring to \cite{zhao2019bearing}, $\lambda_{\min}(\mathcal{B}_{ff})$ can be interpreted as a measure of the degree of uniqueness of the target formation $(\mathcal{G}, p^*)$.
\end{Remark}

%The following corollary is another consequence of Theorem \ref{bbmain}.

%\begin{corollary}\label{choi}
%The \textit{bearing formation error} $e_a$ converges to zero if the exogenous disturbance $f$ converges to zero as time goes to infinity.

%\begin{IEEEproof}
%Consider the Lyapunov candidate function $V_a$ in (\ref{cull}). From the last inequality in (\ref{singu})
%\begin{equation}\label{hannnb}
%\dot{V}_a \le -\delta{V_a} + \gamma^2||f||^2,
%\end{equation}
%$V_a(t)$ is given by
%\begin{equation}\label{jeongy}
%\begin{split}
%V_a(t) &\leq {e}^{-\delta (t-t_0)}V_a(t_0) + \int_{t_0}^{t}{e}^{-\delta(t-\tau)}\gamma^2||f(\tau)||^2d\tau \\
%& \leq {e}^{-\delta (t-t_0)}V_a(t_0)  +\frac{1-{e}^{-\delta(t-t_0)}}{\delta}\sup_{t_0\leq \tau \leq t} \gamma^{2}||f(\tau)||^2.
%\end{split}
%\end{equation}
%Therefore, if the exogenous disturbance $f(t)$ converges to zero as time goes to infinity, it follows that $\lim_{t \rightarrow \infty}V_a(t) = 0$ which means that the \textit{bearing formation error} $e_a$ converges to zero.
%\end{IEEEproof}
%\end{corollary}

%%%%%%%%%%%%%%%%%%%%%%%%%%%%%%%%%%%%%%%%%%%%%%%

\section{Bearing-based network localization with exogenous disturbances}\label{sec4}
%In this section, we analyze the stability of the bearing-based network localization system in the presence of the exogenous disturbances. 

\subsection{Problem statement}
Consider a stationary localization network $(\mathcal{G}, p)$ of node set $\mathcal{V}$ consists of leader set $\mathcal{V}_L$ and follower set $\mathcal{V}_F$ where $\mathcal{V} = \mathcal{V}_L \cup \mathcal{V}_F$, i.e., $|\mathcal{V}_L|=n_l, |\mathcal{V}_F|=n_f, n = n_l + n_f$. In network localization, the positions of the $n_l$ leader nodes are given, while the positions of the $n_f$ follower nodes will be estimated by bearing-based localization protocol. We assume that each agent can sense a global orientation. Therefore, the measured relative bearing $g_{ij}$ can be expressed with respect to a common global orientation. Also, each agent can transmit the own estimated position $\hat{p}_i$ to its neighboring agents.

The bearing-based network localization problem in this paper is stated as follows:

\begin{Problem}{(\textit{Bearing-based Network Localization})}:\label{probb}
For a localization network $(\mathcal{G}, p)$, where $p = [p_l^T,\ p_f^T]^T$, assume that the relative bearing $g_{ij}$ and the positions of the leader nodes $p_l$ are given. Then, the bearing-based network localization problem is to estimate and determine the positions of the follower nodes $\hat{p}_f$. 
\end{Problem}
From \cite{zhao2016localizability}, Problem \ref{probb} can be converted to the following least-squares problem
\begin{equation}\label{least}
\begin{split}
\min_{\hat{p} \in \mathbb{R}^{dn}}J(\hat{p}) = \frac{1}{2}\sum_{i \in \mathcal{V}}&\sum_{j \in \mathcal{N}_i}||P_{g_{ij}}(\hat{p}_i - \hat{p}_j)||^2,\\
\text{subject to} \quad  \hat{p}_i &= p_i,\quad i \in \mathcal{V}_L,
\end{split}
\end{equation}
where $\hat{p}_i$ is the estimated position of agent $i$, and the function $J(\hat{p})$ can be expressed as 
\begin{equation}\label{quadratic}
J(\hat{p}) = \hat{p}^T\mathcal{B}(\mathcal{G}(p))\hat{p},
\end{equation}
where $\mathcal{B}(\mathcal{G}(p))$ is the bearing Laplacian matrix for $(\mathcal{G}, p)$ and $\hat{p} = [{p}_l^T,\ \hat{p}_f^T]^T$ is the estimated position vector set. 
The localization network is assumed to be \textit{bearing localizable} as noted in Definition \ref{localize} to localize in a true network location. 

%We consider Problem \ref{probb} in the presence of the exogenous disturbances in Section \ref{jaja}. In the real-world, the localization network system may face the various source of exogenous disturbances, e.g., wireless communication error, bearing measurement error, then there may exist the \textit{bearing localization error}. For this situation, therefore, it is worth analyzing on the impact of the exogenous disturbances to the localization error.

\subsection{Stability analysis for bearing-based network localization system}\label{jaja}
From the partitioned bearing Laplacian matrix for $(\mathcal{G}, p)$, we state the following lemma for later stability analysis.
\begin{Lemma}{\cite{zhao2016localizability}}\label{jjin}
For the \textit{bearing localizable} network $(\mathcal{G}, p)$, $\mathcal{B}_{ff}$ is positive definite and satisfies
\begin{equation}\label{sujin}
\mathcal{B}_{fl}p_{l} + \mathcal{B}_{ff}p_f = 0.
\end{equation}
\end{Lemma}

Based on the distributed bearing-based localization protocol that ensures global localization\cite{zhao2016localizability}, we consider the bearing-based network localization system with the exogenous disturbances as
\begin{equation}\label{loc}
\dot{\hat{p}}_i(t) = -\sum_{j \in \mathcal{N}_i}P_{g_{ij}}(\hat{p}_i(t) - \hat{p}_j(t)) + z_i(t), i \in \mathcal{V}_F
\end{equation}
where $z_i(t) \in \mathbb{R}^d$ denotes the time-varying exogenous disturbance of agent $i$. 

From (\ref{loc}), the overall bearing-based localization protocol for the follower nodes is derived as
\begin{equation}\label{loc2}
\dot{\hat{p}}_f = -\mathcal{B}_{ff}\hat{p}_f - \mathcal{B}_{fl}p_l + z,
%\dot{\hat{p}} = -\nabla_{\hat{p}_f}\tilde{J}(\hat{p}_f) = -\mathcal{B}_{ff}\hat{p}_f - \mathcal{B}_{fa}p_a
\end{equation}
where $\hat{p}_f, z \in \mathbb{R}^{dn_f}$ denote the estimated position and the time-varying exogenous disturbances of the follower nodes.

\begin{Assumption}\label{ala}
The time-varying exogenous disturbances are uniformly upper-bounded, i.e., $||z(t)|| \le \sqrt{{\sum_{i= n_l + 1}^n}v_i^2} = \mathcal{F}$, where $\mathcal{F}$ is a small positive constant.  
\end{Assumption}

%Under Assumption \ref{ala}, note that the upper-boundedness of the exogenous disturbance is given as $||z|| \le \sqrt{{\sum_{i=n_l + 1}^n}v_i^2}$.

Since the positions of the leader nodes are given, we define the \textit{bearing localization error} as
\begin{equation}\label{localerror}
e_c := \hat{p}_f - p_f.
\end{equation}
From (\ref{localerror}), the error dynamics is given as 
\begin{equation}\label{kyung}
\dot{e}_c = \dot{\hat{p}}_f.
\end{equation}

The following theorem shows a global exponential ultimate boundedness of the \textit{bearing localization error} $e_c$.
\begin{Theorem}\label{yanam}
Consider the bearing-based network localization system (\ref{loc}) in the presence of the exogenous disturbances. Under Assumption \ref{ala}, the \textit{bearing localization error} $e_c$ is globally exponentially convergent and ultimately bounded for the bounded exogenous disturbances. Furthermore, the \textit{bearing localization error} $e_c$ globally exponentially converges to the bound set $\mathcal{S}_c$
\begin{equation}\label{uncom}
\mathcal{S}_c = \left\{ e_c : ||e_c||^2 \le  \frac{\gamma^2\mathcal{F}^2}{\lambda_{\min}({\mathcal{B}}_{ff}) - {\gamma^{-2}}/{4} - {\delta}/{2}} \right\},
\end{equation}
where $\lambda_{\min}(\mathcal{B}_{ff})$ is the smallest positive eigenvalue of $\mathcal{B}_{ff}$ and $\gamma, \delta$ are positive constants satisfying $\lambda_{\min}({\mathcal{B}}_{ff})-\gamma^{-2}/4>\delta/2$.
\begin{proof}
We consider the following Lyapunov candidate function
\begin{equation}\label{lya}
V_c = \frac{1}{2}||e_c||^2. 
\end{equation}
From (\ref{lya}), the time derivative of $V_c$ is derived as
\begin{equation}\label{bulam}
\begin{split}
\dot{V}_c &= e_c^T\dot{e_c} \\
&= e_c^T\dot{\hat{p}}_f \\
&= -e_c^T\mathcal{B}_{ff}\hat{p}_f-e_c^T\mathcal{B}_{fl}p_l+e_c^Tz \\
&= -e_c^T\mathcal{B}_{ff}e_c -e_c^T\mathcal{B}_{ff}p_f - e_c^T\mathcal{B}_{fl}p_l + e_c^Tz \\
&= -e_c^T\mathcal{B}_{ff}e_c + e_c^Tz,
\end{split}
\end{equation}
where we have used the result (\ref{sujin}) to get the last equation.
Since $\mathcal{B}_{ff}$ is positive definite for $(\mathcal{G}, p)$ from Lemma \ref{jjin}, (\ref{bulam}) can be written as
\begin{equation}\label{suyn}
\begin{split}
\dot{V}_c &\leq -\lambda_{\min}({\mathcal{B}}_{ff})||e_c||^2 + e_c^Tz \\
&\leq -\lambda_{\min}({\mathcal{B}}_{ff})||e_c||^2 + \frac{\gamma^{-2}}{4}||e_c||^2 + \gamma^2 ||z||^2 \\
&\leq -(\lambda_{\min}({\mathcal{B}}_{ff}) - \frac{\gamma^{-2}}{4})||e_c||^2 + \gamma^2 ||z||^2,
\end{split}
\end{equation}
where $\lambda_{\min}({\mathcal{B}}_{ff})$ is the smallest positive eigenvalue of $\mathcal{B}_{ff}$ and we have used Lemma \ref{young} to decompose the term $e_c^Tz$. Also, (\ref{suyn}) can be written as
\begin{equation}\label{horan}
\dot{V}_c \le -(\lambda_{\min}({\mathcal{B}}_{ff}) - \frac{\gamma^{-2}}{4})||e_c||^2 + \gamma^2\mathcal{F}^2,
\end{equation}
where $||z||$ is upper-bounded as $\mathcal{F}$ under Assumption \ref{ala}. For a positive constant $\delta > 0$, by choosing $\gamma$ satisfying
\begin{equation}\label{sanna}
\lambda_{\min}({\mathcal{B}}_{ff})-\frac{\gamma^{-2}}{4}>\frac{\delta}{2},
\end{equation}
the negative value of $-(\lambda_{\min}({\mathcal{B}}_{ff}) - {\gamma^{-2}}/{4})||e_c||^2$ is guaranteed. Then, it follows that the \textit{bearing localization error} $e_c$ is ultimately bounded from Lemma \ref{haeb}. Also, (\ref{horan}) can be written as
\begin{equation}\label{jocheol}
\begin{split}
\dot{V}_c \le -\delta{V}_c -(\lambda_{\min}({\mathcal{B}}_{ff}) - \frac{\gamma^{-2}}{4} - \frac{\delta}{2})||e_c||^2 + \gamma^2\mathcal{F}^2.
\end{split}
\end{equation}
Therefore, $\dot{V}_c \leq -\delta{V}_c$ if
\begin{equation}\label{songhoon}
||e_c||^2 > \frac{\gamma^2{\mathcal{F}^2}}{\lambda_{\min}({\mathcal{B}}_{ff}) - {\gamma^{-2}}/{4} - {\delta}/{2}}.
\end{equation}
As a consequence, from (\ref{songhoon}), it follows that the \textit{bearing localization error} $e_c$ is globally exponentially convergent and ultimately bounded to the bounded set $\mathcal{S}_c$ in (\ref{uncom}).
\end{proof}
\end{Theorem}
By calculating the computable set $\mathcal{S}_c$ in (\ref{uncom}), a useful information for a system design can be obtained by following the similar procedure as in Remark \ref{yeun}-\ref{coro}.

%The following corollary is another consequence of Theorem \ref{yanam}.
%\begin{corollary}
%The \textit{bearing localization error} $e_c$ converges to zero if the exogenous disturbance $z$ converges to zero as time goes to infinity.
%\begin{IEEEproof}
%We follow the similar proof to Corollary \ref{choi}.
%\end{IEEEproof}
%\end{corollary}

%%%%%%%%%%%%%%%%%%%%%%%%%%%%%%%%%%%%%%%%%%%

\begin{figure*}[t!]
\centering

\subfigure[Trajectory of the leaderless formation control system. The initial and final positions are denoted by circles and squares, respectively.]{
\includegraphics[width=0.22\textwidth]{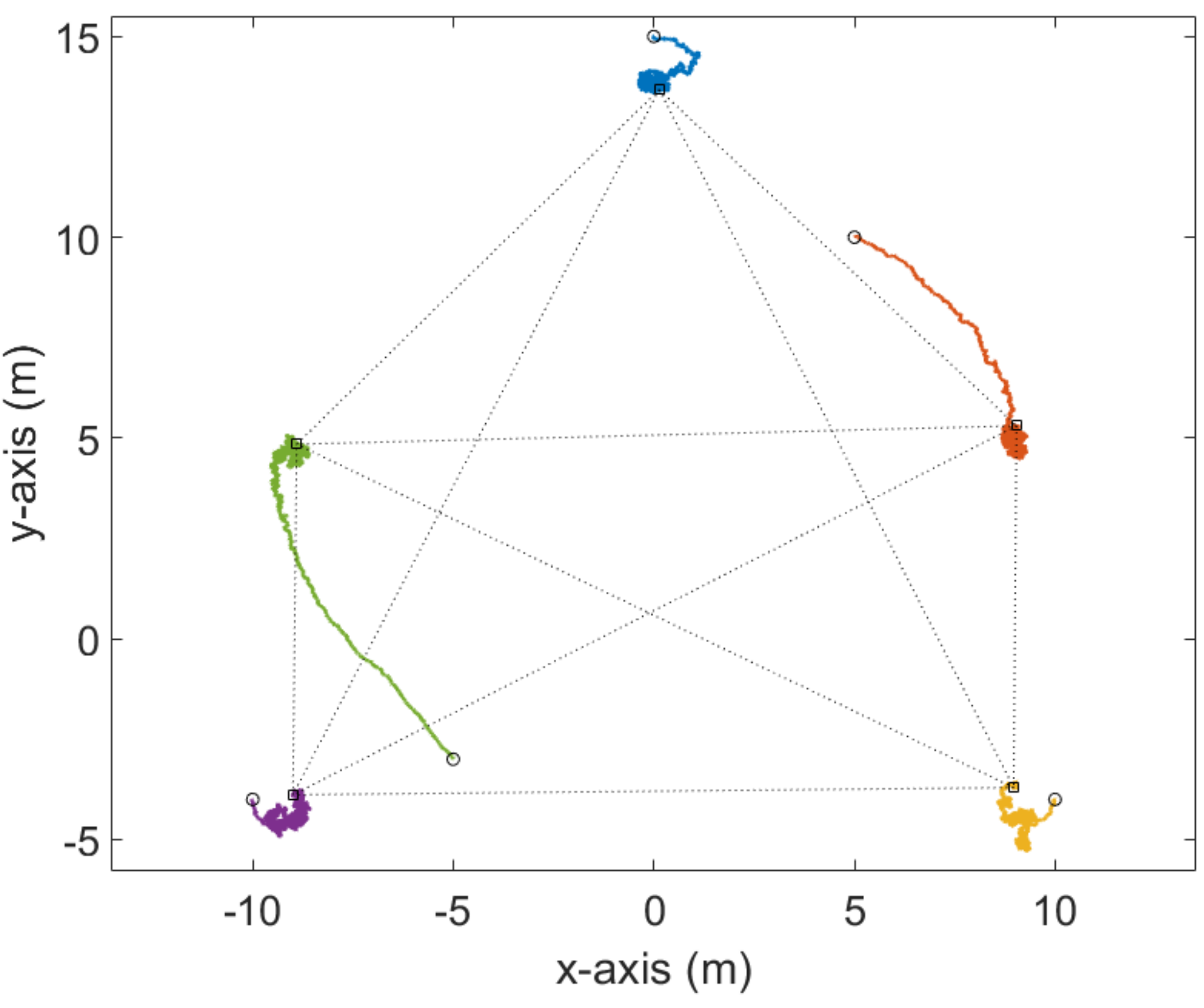}
\label{figure:bearere}
}
\subfigure[Convergence of the bearing formation error $||e_a||$.]{
\includegraphics[width=0.22\textwidth]{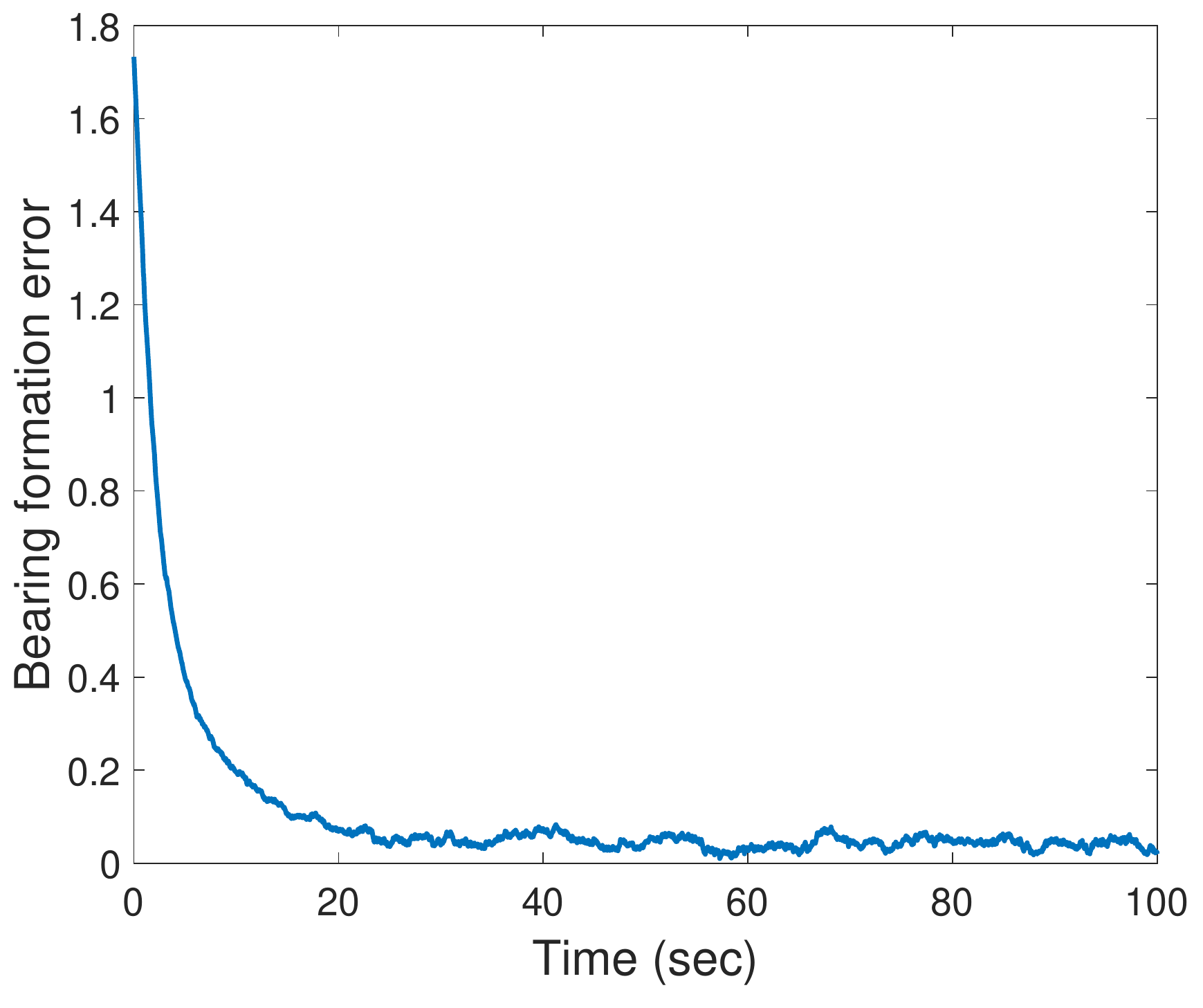}
\label{figure:bearerrrr}
}
\subfigure[Trajectory of the leader-fixed follower formation control system. Two leader agents are denoted by cross. The initial and final positions are denoted by circles and squares, respectively.]{
\includegraphics[width=0.22\textwidth]{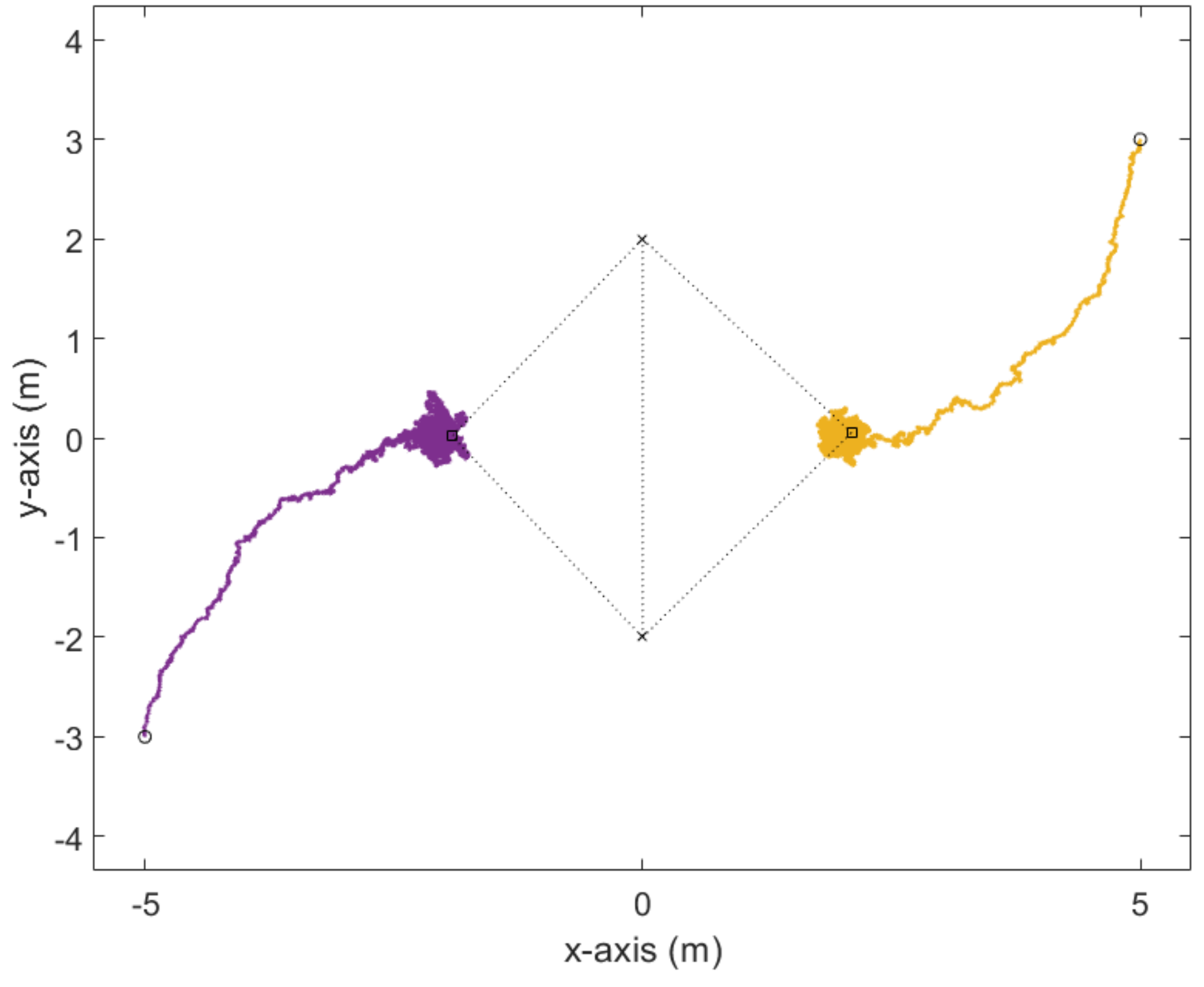}
\label{figure:beartrajj}
}
\subfigure[Convergence of the bearing formation error $||e_b||$.]{
\includegraphics[width=0.215\textwidth]{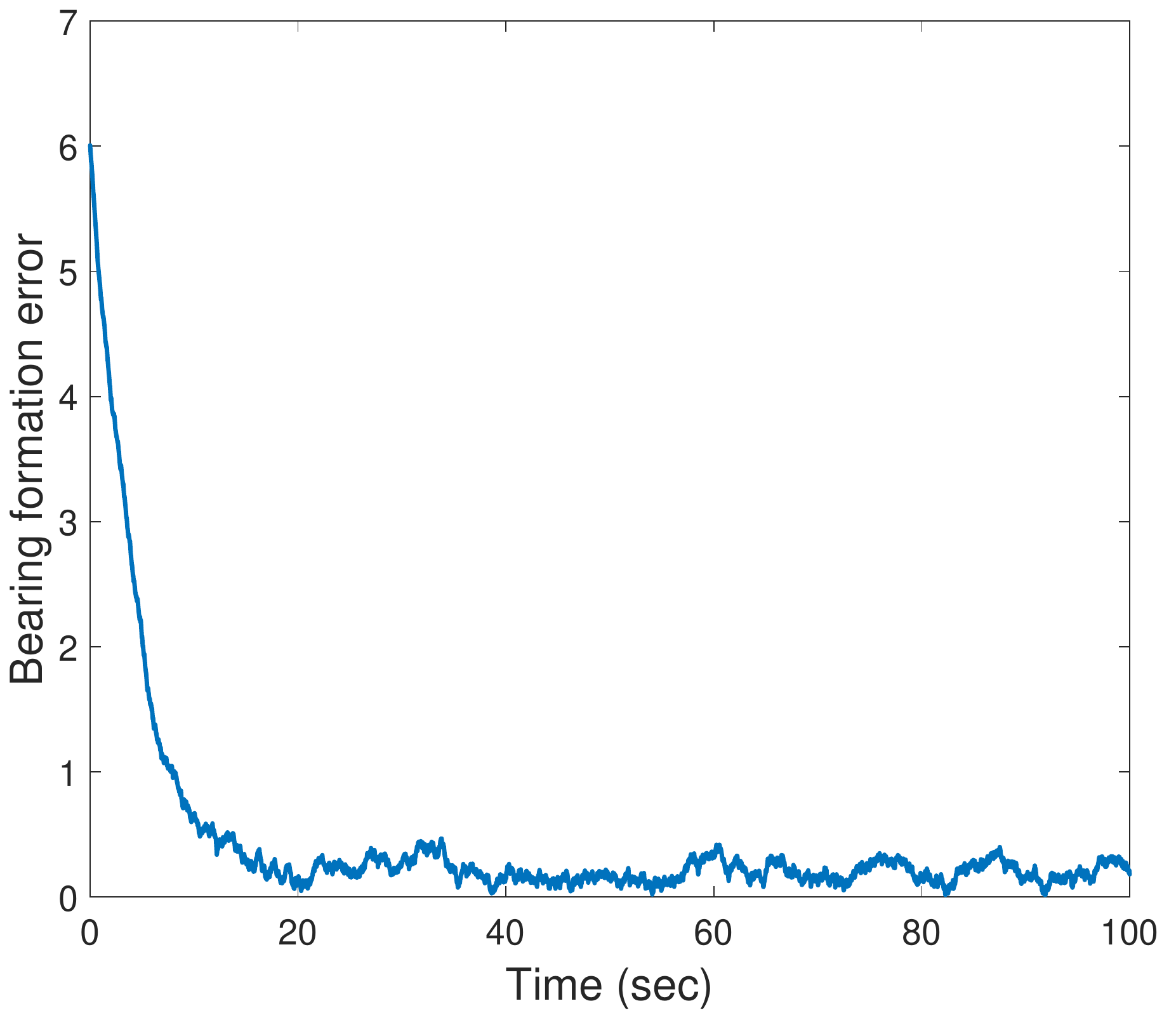}
\label{figure:scalee}
}
\caption{Simulation examples for the bearing-based leader-fixed follower and leaderless formation control systems.}
\label{figure:bearforrrr}
\end{figure*}

\begin{figure*}[t!]
\centering

\subfigure[Stationary bearing localizable network $(\mathcal{G}, p)$. The leader and follower agents are denoted by cross and stars, respectively.]{
\includegraphics[width=0.26\textwidth]{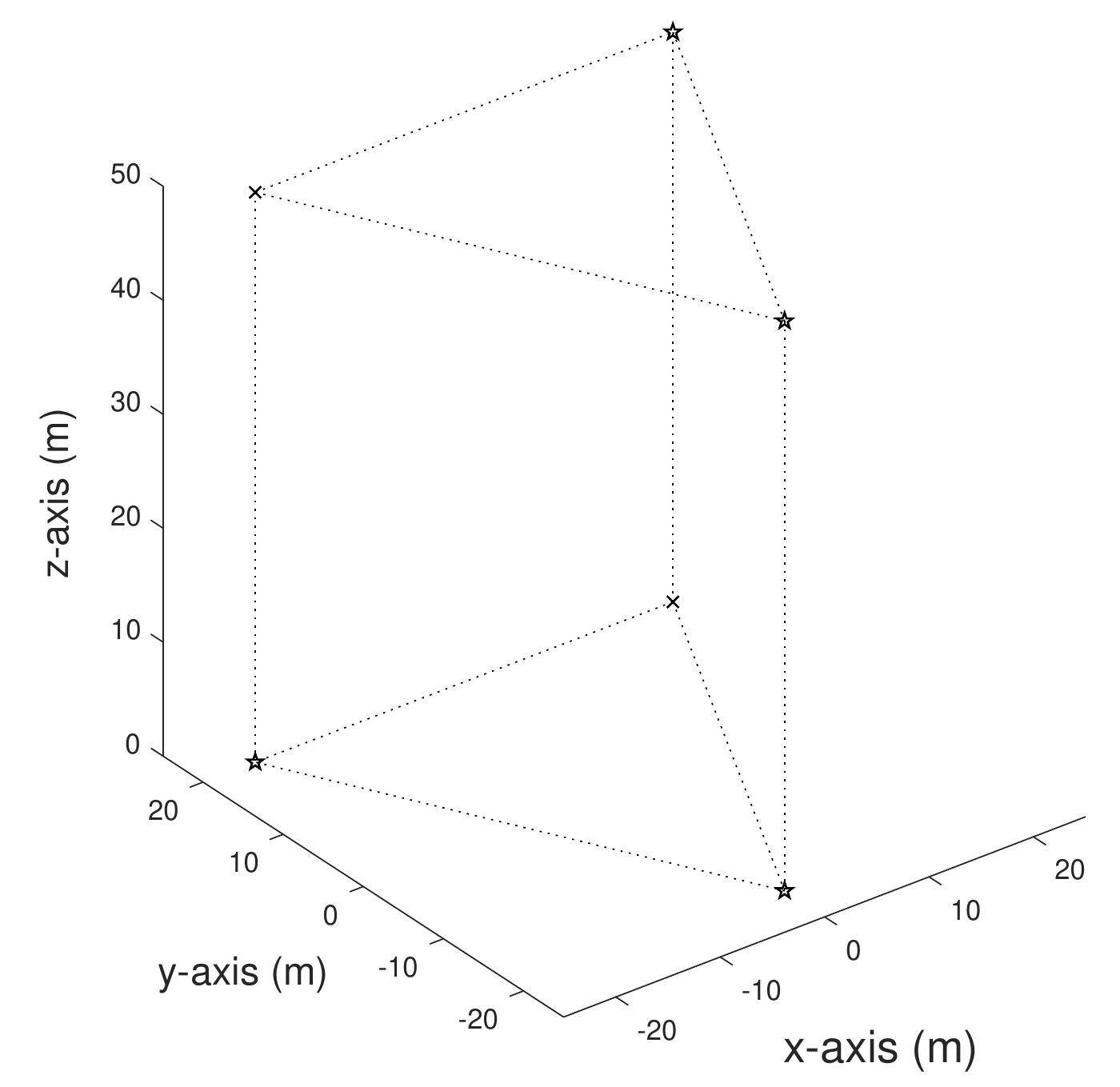}
\label{figure:beartraj}
}
\subfigure[Position estimation. The initial and final estimated positions of the follower agents are denoted by circles and squares, respectively. ]{
\includegraphics[width=0.26\textwidth]{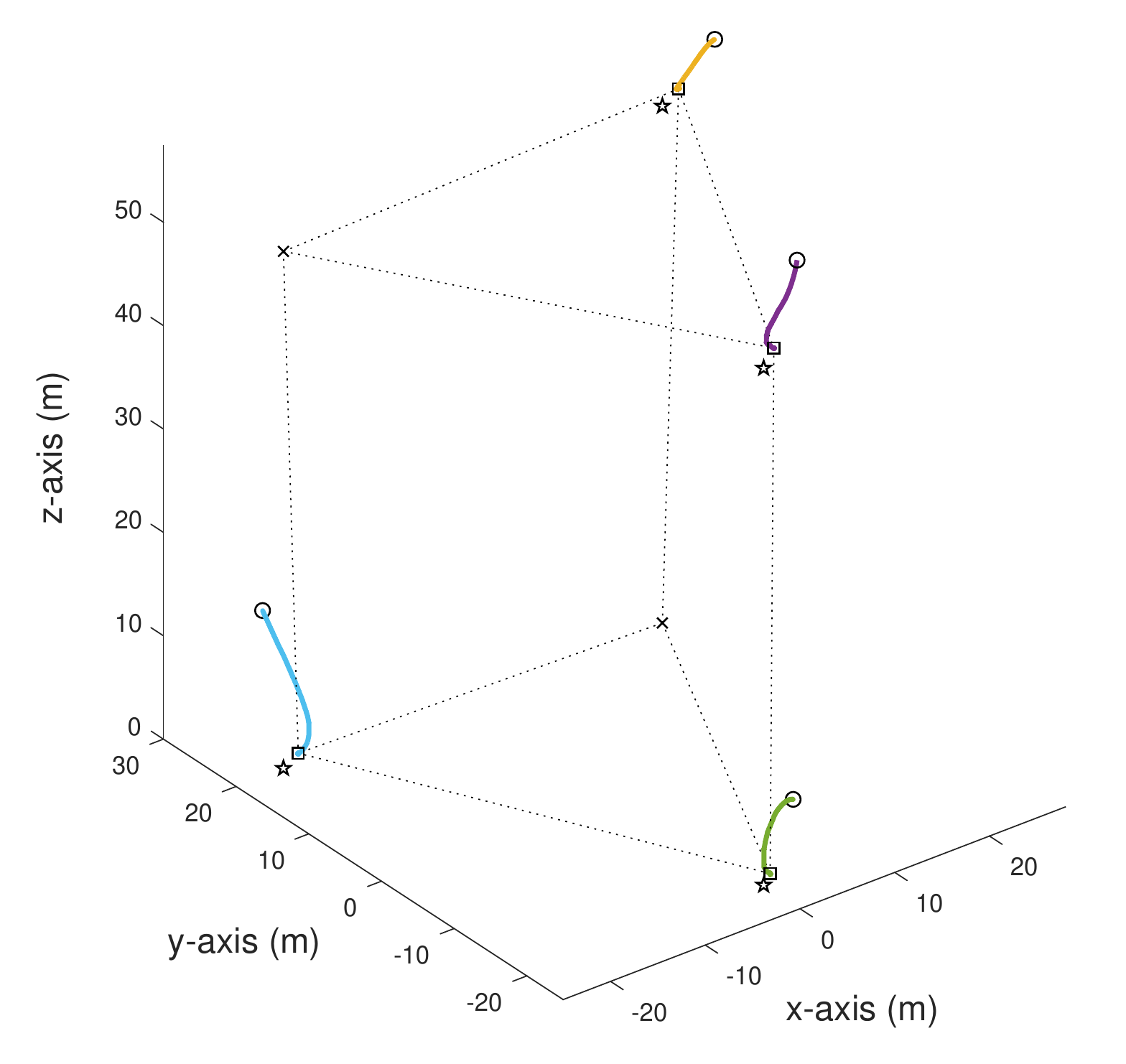}
\label{figure:scale}
}
\subfigure[Convergence of the bearing localization error $||e_c||$.]{
\includegraphics[width=0.29\textwidth]{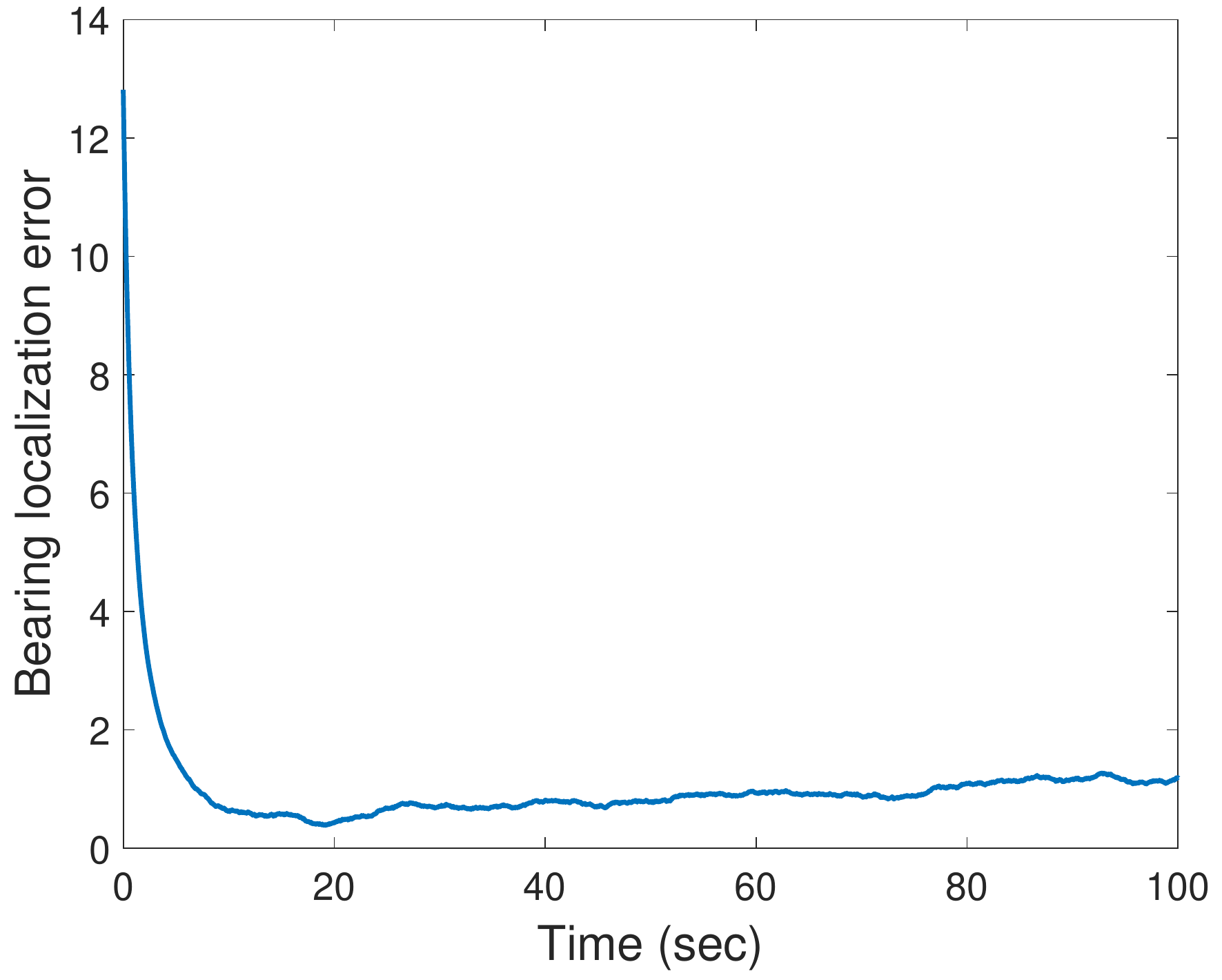}
\label{figure:bearerr}
}
\caption{Simulation examples for the bearing-based network localization system.}
\label{figure:tetra}
\end{figure*}

\section{Simulation examples}\label{sec5}
In this section, simulation examples are provided to validate the proposed stability analysis. The time-varying exogenous disturbances are given as a uniformly distributed random number.

\subsection{Formation control systems with exogenous disturbances}
We first show simulation examples for the bearing-based formation control systems in two-dimensional space, as shown in Fig. \ref{figure:bearforrrr}.
For the leaderless formation control system with the time-varying exogenous disturbances, we consider the infinitesimally bearing rigid formation that contains five agents and ten edges. As shown in Fig. \ref{figure:bearere}-\ref{figure:bearerrrr}, the distributed formation control protocol (\ref{songho}) makes that the formation system converges close to the target formation shape. Also, the bearing formation error $e_a$ converges and upper-bounded as we expected.
For the leader-fixed follower formation control system with the time-varying exogenous disturbances, we consider the bearing localizable formation that contains two leader agents, two follower agents, and five edges. As shown in Fig. \ref{figure:beartrajj}-\ref{figure:scalee}, the distributed formation control protocol (\ref{songgho}) makes that the formation system converges close to the target formation configuration. Also, the bearing formation error $e_b$ converges and upper-bounded.

\subsection{Network localization system with exogenous disturbances}
We next show simulation examples for the bearing-based network localization system with the time-varying exogenous disturbances in three-dimensional space. For the stationary bearing localizable network that contains two leader agents and four follower agents, as shown in Fig. \ref{figure:beartraj}, the network localization protocol (\ref{loc}) estimates the positions of the follower agents. Then, the final estimation $\hat{p}$ is close to the true network position $p$, and the bearing localization error $e_c$ converges and upper-bounded, as shown in Fig. \ref{figure:scale}-\ref{figure:bearerr}.

\section{Conclusion}\label{sec6}
This work studied the undirected bearing-based formation control and network localization systems with the exogenous disturbances in arbitrary dimensional space. In addition to investigating the stability analysis, the explicit upper-bound sets of bearing formation and network localization errors were computed. To be specific, for the leaderless bearing-based formation control system, we showed the local asymptotic bound of the \textit{bearing formation error} $e_a$ with the exogenous disturbances. However, the bound set $\mathcal{S}_a$ was incomputable since the set $\mathcal{S}_a$ consists of the state-dependent matrix $RR^T$. Therefore, only some information could be obtained by computing $\mathcal{S}_a$. To resolve this issue, fixed-leader agents were considered for the leader-fixed follower formation control system. Based on the conventional bearing-only formation control law, the global asymptotic bound of the $\textit{bearing formation error}$ $e_b$ was proved. Also, unlike the leaderless formation control system, the precise bound set $\mathcal{S}_b$ was computable since the existence of the fixed-leader agents makes that the Lyapunov stability analysis is governed by the bearing Laplacian matrix, which is a constant matrix. Therefore, useful information for a system design could be obtained by computing $\mathcal{S}_b$, as stated in Remark \ref{yeun}-\ref{coro}. Lastly, we analyzed the bearing-based network localization system with the exogenous disturbances. Based on the conventional bearing-based localization protocol, a stability of the network localization system was investigated, and the global exponential bound set of the \textit{bearing localization error} $e_c$ was calculated. Likewise, in Section \ref{sec33}, a useful information can be obtained by computing $\mathcal{S}_c$.

The underlying graph in this paper is assumed to be undirected. Therefore, our possible future works would include robust stability analysis on bearing-based formation control and network localization systems for a directed network.
% or


%\appendices

%\section{}

%\section*{Acknowledgment}
%This work was supported by the National Research Foundation (NRF) of Korea under the grant NRF-2017R1A2B3007034.

\ifCLASSOPTIONcaptionsoff
  \newpage
\fi
\bibliographystyle{IEEEtran}
\bibliography{IEEEabrv,BAE.bib}



%\bibliographystyle{IEEEtran}


%\end{IEEEbiography}

\end{document}